\documentclass[%
 reprint,
superscriptaddress,
nofootinbib,
amsmath,amssymb,
 aps,
 prl,
floatfix,
longbibliography
]{revtex4-2}
\usepackage[colorlinks=true,citecolor=blue,filecolor=blue,linkcolor=blue,urlcolor=blue,pdftex]{hyperref}

\usepackage{graphicx}% Include figure files
\usepackage{dcolumn}% Align table columns on the decimal point
\usepackage{bm}% bold math
\usepackage{siunitx}
\usepackage{booktabs}
\usepackage{xspace}
\usepackage[usenames,dvipsnames]{color}
\usepackage{booktabs,graphicx,mathrsfs,verbatim,amsmath,soul}
\usepackage{xspace} 
\usepackage{upgreek}
\usepackage{textcomp}
\usepackage{gensymb}
\usepackage{tabularx}
\usepackage{xcolor}
\usepackage{multirow}
\usepackage{tabularx}
\usepackage{lineno}
\usepackage[colorlinks=true,citecolor=blue,filecolor=blue,linkcolor=blue,urlcolor=blue,pdftex]{hyperref}
\usepackage{orcidlink} % For ORCID links in author list
\usepackage{currfile} % To be able to list current file name in title
\usepackage[separate-uncertainty=true]{siunitx}
\DeclareSIUnit\gevm{\GeV\per\clight\squared}

\DeclareSIUnit{\events}{events}
\DeclareSIUnit{\ton}{t}
\DeclareSIUnit{\tonwhole}{tonne}
\DeclareSIUnit{\year}{y}
\DeclareSIUnit{\yearwhole}{year}
\DeclareSIUnit{\eventspertonyear}{\events\per\ton\per\year}
\DeclareSIUnit{\eventspertonyearwhole}{\events\per\tonwhole\per\yearwhole}

\newcommand{\itsec}[1]{{\it #1 }--- }
\newcommand{\cevns}{{\text{CE}\ensuremath{\nu}\text{NS}}\xspace}
\newcommand{\beight}{\ensuremath{{}^8\mathrm{B} }\xspace}
\newcommand{\gevcsq}{\ensuremath{\mathrm{GeV}/c^2}}

\newcommand{\gevm}{\gevcsq}

\newcommand{\keV}{\ensuremath{\mathrm{keV}}\xspace}

\begin{document}

\preprint{APS/123-QED}
\title{First Search for Light Dark Matter in the Neutrino Fog with XENONnT}
% Autogenerated on 10 Sep, 2024 at 14:10:23
% Generated with make-authorlist.py

% You could put the following into an 'affiliation.tex' file and 
% \include it in your main text

\newcommand{\bologna}{\affiliation{Department of Physics and Astronomy, University of Bologna and INFN-Bologna, 40126 Bologna, Italy}}
\newcommand{\chicago}{\affiliation{Department of Physics, Enrico Fermi Institute \& Kavli Institute for Cosmological Physics, University of Chicago, Chicago, IL 60637, USA}}
\newcommand{\coimbra}{\affiliation{LIBPhys, Department of Physics, University of Coimbra, 3004-516 Coimbra, Portugal}}
\newcommand{\columbia}{\affiliation{Physics Department, Columbia University, New York, NY 10027, USA}}
\newcommand{\lngs}{\affiliation{INFN-Laboratori Nazionali del Gran Sasso and Gran Sasso Science Institute, 67100 L'Aquila, Italy}}
\newcommand{\mainz}{\affiliation{Institut f\"ur Physik \& Exzellenzcluster PRISMA$^{+}$, Johannes Gutenberg-Universit\"at Mainz, 55099 Mainz, Germany}}
\newcommand{\mpik}{\affiliation{Max-Planck-Institut f\"ur Kernphysik, 69117 Heidelberg, Germany}}
\newcommand{\munster}{\affiliation{Institut f\"ur Kernphysik, University of M\"unster, 48149 M\"unster, Germany}}
\newcommand{\nikhef}{\affiliation{Nikhef and the University of Amsterdam, Science Park, 1098XG Amsterdam, Netherlands}}
\newcommand{\nyuad}{\affiliation{New York University Abu Dhabi - Center for Astro, Particle and Planetary Physics, Abu Dhabi, United Arab Emirates}}
\newcommand{\purdue}{\affiliation{Department of Physics and Astronomy, Purdue University, West Lafayette, IN 47907, USA}}
\newcommand{\rice}{\affiliation{Department of Physics and Astronomy, Rice University, Houston, TX 77005, USA}}
\newcommand{\stockholm}{\affiliation{Oskar Klein Centre, Department of Physics, Stockholm University, AlbaNova, Stockholm SE-10691, Sweden}}
\newcommand{\subatech}{\affiliation{SUBATECH, IMT Atlantique, CNRS/IN2P3, Nantes Universit\'e, Nantes 44307, France}}
\newcommand{\torino}{\affiliation{INAF-Astrophysical Observatory of Torino, Department of Physics, University  of  Torino and  INFN-Torino,  10125  Torino,  Italy}}
\newcommand{\ucsd}{\affiliation{Department of Physics, University of California San Diego, La Jolla, CA 92093, USA}}
\newcommand{\wis}{\affiliation{Department of Particle Physics and Astrophysics, Weizmann Institute of Science, Rehovot 7610001, Israel}}
\newcommand{\zurich}{\affiliation{Physik-Institut, University of Z\"urich, 8057  Z\"urich, Switzerland}}
\newcommand{\paris}{\affiliation{LPNHE, Sorbonne Universit\'{e}, CNRS/IN2P3, 75005 Paris, France}}
\newcommand{\freiburg}{\affiliation{Physikalisches Institut, Universit\"at Freiburg, 79104 Freiburg, Germany}}
\newcommand{\napels}{\affiliation{Department of Physics ``Ettore Pancini'', University of Napoli and INFN-Napoli, 80126 Napoli, Italy}}
\newcommand{\nagoya}{\affiliation{Kobayashi-Maskawa Institute for the Origin of Particles and the Universe, and Institute for Space-Earth Environmental Research, Nagoya University, Furo-cho, Chikusa-ku, Nagoya, Aichi 464-8602, Japan}}
\newcommand{\laquila}{\affiliation{Department of Physics and Chemistry, University of L'Aquila, 67100 L'Aquila, Italy}}
\newcommand{\tokyo}{\affiliation{Kamioka Observatory, Institute for Cosmic Ray Research, and Kavli Institute for the Physics and Mathematics of the Universe (WPI), University of Tokyo, Higashi-Mozumi, Kamioka, Hida, Gifu 506-1205, Japan}}
\newcommand{\kobe}{\affiliation{Department of Physics, Kobe University, Kobe, Hyogo 657-8501, Japan}}
\newcommand{\kit}{\affiliation{Institute for Astroparticle Physics, Karlsruhe Institute of Technology, 76021 Karlsruhe, Germany}}
\newcommand{\tsinghua}{\affiliation{Department of Physics \& Center for High Energy Physics, Tsinghua University, Beijing 100084, P.R. China}}
\newcommand{\ferrara}{\affiliation{INFN-Ferrara and Dip. di Fisica e Scienze della Terra, Universit\`a di Ferrara, 44122 Ferrara, Italy}}
\newcommand{\groningen}{\affiliation{Nikhef and the University of Groningen, Van Swinderen Institute, 9747AG Groningen, Netherlands}}
\newcommand{\westlake}{\affiliation{Department of Physics, School of Science, Westlake University, Hangzhou 310030, P.R. China}}
\newcommand{\shenzhen}{\affiliation{School of Science and Engineering, The Chinese University of Hong Kong, Shenzhen, Guangdong, 518172, P.R. China}}
\newcommand{\coimbrapoli}{\affiliation{Coimbra Polytechnic - ISEC, 3030-199 Coimbra, Portugal}}
\newcommand{\uniheidelberg}{\affiliation{Physikalisches Institut, Universit\"at Heidelberg, Heidelberg, Germany}}
\newcommand{\roma}{\affiliation{INFN-Roma Tre, 00146 Roma, Italy}}
\newcommand{\bucknell}{\affiliation{Department of Physics \& Astronomy, Bucknell University, Lewisburg, PA, USA}}

% End of AFFILIATIONS

% You could put the following into an 'authors.tex' file and 
% \include it in your main text

\author{E.~Aprile\,\orcidlink{0000-0001-6595-7098}}\columbia
\author{J.~Aalbers\,\orcidlink{0000-0003-0030-0030}}\groningen
\author{K.~Abe\,\orcidlink{0009-0000-9620-788X}}\tokyo
\author{S.~Ahmed Maouloud\,\orcidlink{0000-0002-0844-4576}}\paris
\author{L.~Althueser\,\orcidlink{0000-0002-5468-4298}}\munster
\author{B.~Andrieu\,\orcidlink{0009-0002-6485-4163}}\paris
\author{E.~Angelino\,\orcidlink{0000-0002-6695-4355}}\torino\lngs
\author{D.~Ant\'on~Martin\,\orcidlink{0000-0001-7725-5552}}\chicago
\author{F.~Arneodo\,\orcidlink{0000-0002-1061-0510}}\nyuad
\author{L.~Baudis\,\orcidlink{0000-0003-4710-1768}}\zurich
% ...
\collaboration{XENON Collaboration}\email[]{xenon@lngs.infn.it}\noaffiliation

%
% End of AUTHORS

\date{\today}

\begin{abstract} We search for dark matter (DM) with a mass [3, 12] \gevcsq ~using an exposure of \SI{3.51}{\tonwhole\times\yearwhole} with the XENONnT experiment. We consider spin-independent DM-nucleon interactions mediated by a heavy or light mediator, spin-dependent DM-neutron interactions, momentum-dependent DM scattering, and mirror DM. Using a lowered energy threshold compared to the previous WIMP search, a blind analysis of [0.5, 5.0] \keV nuclear recoil events reveals no significant signal excess over the background. XENONnT excludes spin-independent DM-nucleon cross sections \SI{>2.5e-45}{cm^2} at 90\% confidence level for 6 \gevcsq ~DM. In the considered mass range, the DM sensitivity approaches the `neutrino fog’, the limitation where neutrinos produce a signal that is indistinguishable from that of light DM-xenon nucleus scattering. 
\end{abstract}

\maketitle

\itsec{Introduction} The XENONnT experiment \cite{XENON:2024wpa}, operated at INFN Laboratori Nazionali del Gran Sasso, is primarily designed to search for weakly interacting massive particles (WIMPs) \cite{Roszkowski:2017nbc,Jungman:1995df,Bertone:2016nfn} with a mass scale of \gevcsq -TeV/$c^2$. By lowering the energy threshold compared to our previous WIMP search \cite{XENON:2023cxc}, we are able to improve the sensitivity to light dark matter (DM). However, as we increase the exposure, the solar \beight neutrino coherent elastic neutrino-nucleus scattering (\cevns) background begins to affect the sensitivity gain. This gradual slowdown in sensitivity improvement for light DM with additional exposure is known as the neutrino fog \cite{OHare:2021utq,Billard:2013qya,Ruppin:2014bra,Akerib:2022ort,Billard:2021uyg}. In this Letter, we report on the first search for light DM with a mass of [3, 12] \gevcsq, achieving DM sensitivities into the neutrino fog.
This search utilizes the same dataset and analysis techniques as in our recently reported search for solar \beight neutrino via \cevns interaction \cite{XENON:2024ijk}.

The XENONnT experiment comprises a central detector, operating as a time projection chamber (TPC), with an active target of 5.9 tonnes of cryogenic liquid xenon, enclosed by a Cherenkov neutron veto (NV) and muon veto \cite{XENON1T:2014eqx} detector to mitigate radiogenic neutrons and cosmogenic muon-induced backgrounds. The operation of XENONnT is supported by various subsystems, including a gas-phase and a liquid-phase xenon circulation and purification system, a radon removal system \cite{Murra:2022mlr}, and a krypton distillation column \cite{XENON:2016bmq}. A particle interaction in the xenon target may produce a nuclear recoil (NR) with the xenon nucleus, or an electronic recoil (ER) with the atomic electrons. Two arrays of photomultiplier tubes (PMTs) at the top and bottom of the TPC collect the light from the scintillation signal (S1) and the ionization signal (S2), converting them to photo-electrons (PE). Digitizers convert the PE waveforms that cross the digitization threshold into PMT hits, which are clustered into peaks by reconstruction software \cite{strax, straxen}.
The detector wall consists of Polytetrafluoroethylene (PTFE) panels to enhance photon gain (g1). The cathode electrode near the bottom of the TPC and the gate electrode below the liquid-gas interface establish a vertical electric field of approximately \SI{23}{V/cm}, which drifts the ionization electrons to the liquid-gas interface with a maximum drift time of \SI{2.2}{ms}. An extraction field is generated by the gate and anode electrodes, leading to secondary scintillation of the extracted electrons. The extraction efficiency and single electron scintillation gain contribute to the electron gain (g2). Further details on XENONnT can be found in Ref. \cite{XENON:2024wpa}.

\itsec{Dataset} This DM search combines data from two Science Runs spanning from May 1, 2021, to December 10, 2021 (SR0, livetime 108.0 days), and from May 19, 2022, to August 8, 2023 (SR1, livetime 208.5 days). Between SR0 and SR1, the radon distillation system was upgraded to operate in both gas xenon (GXe) and liquid xenon (LXe) modes. On July 15, 2022, the liquid level was decreased by 0.2 mm from \SI{5.0\pm0.2}{mm} and the anode voltage was increased from \SI{4.90}{kV} to \SI{4.95}{kV} to mitigate clustered electron emissions \cite{Tomas:2018pny,LUX:2020vbj}, while keeping a constant electron gain. In SR0 (SR1), the average electron lifetime is \SI{14.2}{ms} (\SI{20.1}{ms}), g1 is $0.1515 \pm 0.0014$\,PE/photon ($0.1367 \pm 0.0010$\,PE/photon), and g2 is $16.45 \pm 0.64$\,PE/e$^-$ ($16.85 \pm 0.46$\,PE/e$^-$).

\begin{figure}[h!t]
    \includegraphics[width=1.0\columnwidth]{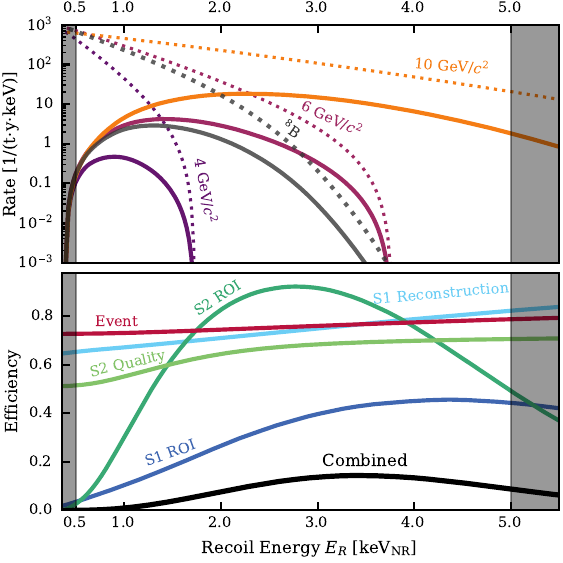}
    \caption{\label{fig:efficiency}
    Top: The SI DM signal spectrum with a cross section of 4.4 $\times$ 10$^{-45}$ cm$^2$ before (dotted) and after (solid) accounting for the signal efficiency. The \beight~ \cevns spectrum (gray) is shown for comparison.
    Bottom: The signal efficiency (black) is a combination of individual efficiencies of S1 ROI (blue), S1 reconstruction (light blue), S2 ROI (dark green), S2 quality selection (light green), and event efficiency (red). These efficiencies are evaluated per science run and weighted by livetime.
    }
\end{figure}

\itsec{Signal} 
We consider DM models involving nuclear recoils from the elastic scattering of DM particles on xenon nuclei, focusing on a mass ranging from 3 to 12 \gevcsq. This mass range is favored by various DM models such as WIMP \cite{Roszkowski:2017nbc,Jungman:1995df,Bertone:2016nfn}, asymmetric DM \cite{Petraki:2013wwa,Kaplan:2009ag}, and self-interacting DM \cite{Spergel:1999mh,Tulin:2017ara,Zurek:2008qg,Fornengo:2011sz}. We examine spin-independent (SI) DM-nucleon \cite{Lewin:1995rx} and spin-dependent (SD) DM-neutron  \cite{Menendez:2012tm} interactions mediated by a heavy mediator with mass $m_\phi \gg q_0/c \equiv10~\mathrm{MeV}/c^2$, where $q_0$ represents the typical momentum transfer in this search. We also consider DM with SI interaction mediated by a scalar or vector light mediator with mass $m_\phi \ll q_0/c$ (SI-LM) \cite{DelNobile:2015uua, Fornengo:2011sz, XENON:2019gfn}. In this case, the differential rate is proportional to $\sigma m_\phi^4/(q^2/c^2+m_\phi^2)^2\approx \sigma m_\phi{ }^4c^4/q^4$, where $q$ denotes the momentum transfer, and $\sigma$ is the DM-nucleon cross section \cite{PandaX-II:2018xpz, PandaX:2022xqx}. Momentum-dependent DM scattering (MDDM) \cite{Chang:2009yt} is also considered, characterized by a modified SI interaction cross section of the form $\sigma_{\chi N}=\left(q / q_0\right)^{2 n} \sigma_0$, where $n\in\left\{1,2\right\}$ and $\sigma_0$ is the DM-nucleon cross section. Additionally, we search for mirror dark matter, characterized by a differential cross section resembling Rutherford scattering \cite{Foot:2014mia, Clarke:2016eac}.
The recoil energy spectra are evaluated using Ref. \cite{jelle_aalbers_2023_7636982}, and the spectra for SI interaction are shown in Fig. \ref{fig:efficiency} (top). 
For astrophysical models and nuclear form factors, we follow Ref. \cite{XENON:2023cxc} with the exception of mirror DM, where we adopt a DM-mass-dependent Boltzmann velocity $v_0$ as assumed in Ref. \cite{Clarke:2016eac} beyond the Standard Halo Model recommended in Ref. \cite{Baxter:2021pqo}.

The average number of scintillation photons and ionization electrons from the energy of an NR interaction ($E_R$) is quantified by the light yield model $L_y$ and the charge yield model $Q_y$. We model detector responses for low-energy NR events \cite{xenonnt_ybe} with $^{88} \mathrm{Y} \mathrm{Be}$ neutron calibration data \cite{collar_applications_2013} (with a mean $\bar{E}_R \approx 2$ keV per scatter), and fit $L_y$ and $Q_y$ following the methodology described in Ref. \cite{XENON:2024xgd}.
Yields for $E_R< \SI[inter-unit-product=\;]{0.5}{keV}$ are assumed to be zero, and light DM signals with $E_R> \SI[inter-unit-product=\;]{5.0}{keV}$ are neglected. 
For $E_R< \SI[inter-unit-product=\;]{5.0}{keV}$, the $^{88} \mathrm{Y} \mathrm{Be}$ calibrated yield model aligns with the $Q_y$ of the NEST v2.3.6 \cite{szydagis_nest_2011} yield model within systematic uncertainty, while the $L_y$ uncertainties generally encompass the median of the NEST model.
The uncertainties of $L_y$ and $Q_y$ are modeled using shape parameters $t_{\mathrm{Ly}}$ and $t_{\mathrm{Qy}}$, respectively.  $t_{\mathrm{Ly}}$ and $t_{\mathrm{Qy}}$ determine the relative shift of $L_y$ and $Q_y$ from the median towards the upper and lower bound of their 68\% credible intervals \cite{xenonnt_ybe}.

A simulation-based detector response model converts the produced photons and electrons to detector observables \cite{xenonnt_analysis1}. 
Using the depth ($Z$) and radius ($R$) position of the event, the fiducial volume (FV) is defined as $-142.0$ cm $\leq Z \leq -13.0$ cm, $R \leq 60.2 \mathrm{~cm}$ for SR0 and $R \leq 59.6 \mathrm{~cm}$ for SR1.
The region of interest (ROI) is defined as 2 or 3 hits for S1 and [120, 500] PE for S2, corresponding to approximately [4, 16] extracted electrons.
Events reconstructed with $R\leq63.0$ cm in the ROI were blinded before finalizing the analysis. Compared to the previous XENONnT NR search \cite{XENON:2023cxc}, the ROI in this Letter leads to a 17 times higher signal rate expectation for a $\SI[inter-unit-product=\;]{6}{\gevcsq}$  ~SI DM. The overlap of ROI between this Letter and the previously unblinded SR0 WIMP search \cite{XENON:2023cxc} was not used in the data selection development, and is used in this search.

Three types of selection are implemented, following the \beight~\cevns analysis \cite{XENON:2024ijk}.
For the S1 peaks, the reconstruction requires at least two hits from two different PMTs within \SI{50}{ns}, which is referred to as the 2--fold PMT coincidence.
For the S2 peaks, other than the waveform width and PMT distribution quality selections inherited from Ref. \cite{XENON:2023cxc}, an S2 boosted-decision-tree machine is developed as detailed in the background section. 
To distinguish physical events from detector artifacts, we implement selections that consider temporal and spatial correlations with preceding high-energy (HE) events \cite{XENON:2024ijk}, including $\mathrm{S} 2_{\text {pre}} / \Delta t_{\text {pre}}$, defined as the ratio between the S2 peak area of HE events and the time difference from candidate events. Following Ref. \cite{XENON:2023cxc}, a single-scatter selection is applied. A selection based on the single electron (SE) rate within a preceding 5-second time window is implemented to mitigate the impact of single electron pile-up \cite{Tomas:2018pny,LUX:2020vbj}.

\begin{table*}[h!tbp]
\caption{\label{tab:table1}
For each model component, we show the background expectation and best-fit values for the background-only hypothesis, background and SI DM signal hypothesis with a 4, 6, 10 \gevcsq ~mass DM. 
The expectation uncertainties for $^8\mathrm{B}$ \cevns background consider both signal efficiency uncertainty and yield uncertainties while neglecting the subdominant 3\% \beight neutrino flux uncertainty \cite{SNO:2011hxd}. 
All uncertainties in the last four columns are calculated using the 68$\%$ confidence interval constructed by the profile likelihood method.
In addition to the event number in the full ROI, the event number in the most signal-like region is shown in parentheses.
The best-fit and expectation agree within the uncertainties for all components.} 
\begin{ruledtabular}
\newcolumntype{R}{>{\raggedleft\arraybackslash}p{1.5cm}} % Define a custom column type for the numbers
\newcolumntype{L}{>{\raggedright\arraybackslash}p{1.5cm}} % Define a custom column type for the numbers after parenthesis

\begin{tabular}{
  @{}l
  c
  c
  R@{~}L
  R@{~}L
  R@{~}L
  @{}
}
\textbf{Component} & \textbf{Expectation} & \textbf{Background-only} & \multicolumn{2}{c}{\textbf{4 \gevcsq}} & \multicolumn{2}{c}{\textbf{6 \gevcsq}} & \multicolumn{2}{c}{\textbf{10 \gevcsq}} \\
\midrule
SI DM & -- & -- & 3.2 & (1.6) & 0.0 & (0.0) & 0.0 & (0.0) \\
$^8\mathrm{B}$ CE$\nu$NS & $11.9_{-4.2}^{+4.5}$ & $11.4^{+2.0}_{-3.6}$ & 10.2 $\pm 2.7$ &$(2.5 \pm 0.7)$ & $11.4^{+2.7}_{-2.6}$ & $(6.0 \pm 1.4)$ & $11.4^{+2.7}_{-2.6}$ & $(5.4_{-1.2}^{+1.3})$ \\
AC & $25.3 \pm 1.2$ & $25.3_{-1.3}^{+1.1}$ & 25.1 $\pm 1.2$ &$(3.7 \pm 0.1)$ & $25.3 \pm 1.2$ & $(4.1 \pm 0.2)$ & $25.3 \pm 1.2$ & $(3.3 \pm 0.1)$ \\
ER & $0.7 \pm 0.7$ & $0.5^{+0.7}_{-0.6}$ & 0.6 $_{-0.6}^{+0.7}$ &$(0.0 \pm 0.0)$ & $0.5^{+0.6}_{-0.5}$ & $(0.3 \pm 0.3)$ & $0.5_{-0.5}^{+0.7}$ & $(0.4_{-0.4}^{+0.5})$ \\
Neutron & $0.5^{+0.2}_{-0.3}$ & $0.5 \pm 0.3$ & 0.5 $\pm 0.3$ &$(0.1 \pm 0.1)$ & $0.5 \pm 0.3$ & $(0.2 \pm 0.1)$ & $0.5 \pm 0.3$ & $(0.2 \pm 0.1)$ \\
\midrule
Total background & $38.3_{-4.4}^{+4.7}$ & $37.7_{-3.9}^{+2.5}$ & 36.4 $_{-3.0}^{+3.0}$ &$(6.3_{-0.7}^{+0.7})$ & $37.7_{-2.9}^{+3.0}$ & $(10.6_{-1.4}^{+1.5})$ & $37.7_{-2.9}^{+3.0}$ & $(9.2_{-1.3}^{+1.4})$ \\
Observed & -- & 37 & 37 & (10) & 37 & (10) & 37 & (4) \\
\end{tabular}
\end{ruledtabular}
\end{table*}

The signal efficiency consists of the ROI efficiency, the S1 reconstruction efficiency, the S2 quality selection efficiency, and the event efficiency, as illustrated in Fig. \ref{fig:efficiency} (bottom).  Excluding yield model uncertainties, the systematic uncertainty of the combined efficiency is estimated to be 25\%, predominantly due to the uncertainties in the S1 reconstruction efficiency and the S2 quality selection efficiency.

The ROI efficiency is assessed through detector reconstruction simulation and is sensitive to the yield model. For a 6 \gevcsq~ SI DM, the signal efficiency uncertainty due to the yield model alone is estimated to be 34\%, resulting from the propagation of $t_{\mathrm{Ly}}$ and $t_{\mathrm{Qy}}$ to the signal rate computation.

The S1 reconstruction efficiency originates from the two-fold PMT coincidence requirement. We employ a data-driven method \cite{xenonnt_analysis1} that simulates low energy S1 waveforms by sampling a subset of photon waveforms from $^{83\mathrm{m}}$Kr 32.1 keV and 9.4 keV conversion electrons and $^{37}$Ar 2.8 keV electron capture calibration data.  We incorporate a conservative systematic uncertainty of 10\% to account for the systematic uncertainty in the shape of the S1 waveform. The S1 reconstruction efficiency is 65\% (85\%) for S1 signals with 2 (3) detected photons, while the other S1 quality selections achieve acceptances over 99\% and are thus neglected in efficiency evaluation.

\begin{figure*}[htbp]
\centering
\includegraphics[width=1.0\textwidth]{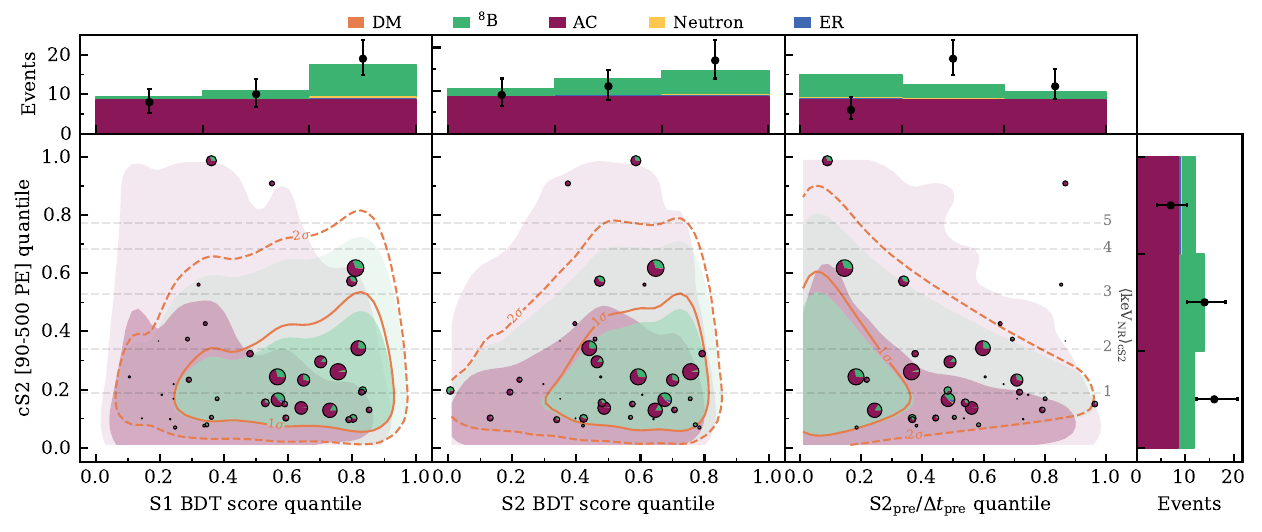}
\caption{\label{fig:analysis_dimension} DM candidate events in the search space, distributed in quantiles of (S1 BDT, cS2) (left), (S2 BDT, cS2) (center) and ($\mathrm{S} 2_{\text{pre}} / \Delta t_{\text{pre}}$, cS2) (right). The quantiles are defined to achieve an equal-probable binning of the AC background. Events after selections are shown as pie charts, with each wedge representing the contribution of the components in the best-fit model assuming a 6 \gevcsq ~DM. The size of the pie charts is proportional to the DM signal model at that position. The contributions of DM, ER, and neutron events in the wedges and histograms are barely visible since the best-fit rates are negligible. The shaded contours show the 1$\sigma$ and 2$\sigma$ distributions of the AC background (purple) and solar \beight ~\cevns background (green). The solid and dashed line contours show the 1$\sigma$ and 2$\sigma$ distribution of a 6 \gevcsq ~DM signal (orange), showing its similarity with  \beight~\cevns. The equivalent nuclear recoil energy based solely on cS2 is indicated by the gray lines. The one-dimensional projection of best-fit and observed data (black) is shown on the side.
}
\end{figure*}

The S2 quality selection efficiency is evaluated by studying the selection survival ratio of evaluation datasets within the ROI, including waveform simulation \cite{fuse}, surface $^{210}$Pb $\beta$ events from PTFE, $^{214}$Pb $\beta$ events from the cathode, $^{37}$Ar \SI{0.27}{keV} electron capture, and NV-tagged low energy $^{241} \mathrm{AmBe}$ neutrons. A systematic uncertainty of 10\% is adopted based on the variation among the evaluation datasets.

The event reconstruction algorithm pairs one S1 to one S2 peak to build an event. The algorithm may fail when S1 and S2 peaks of areas similar to candidate events occur within a time window of $\mathcal{O}(10) ~\mathrm{ms}$. To account for this effect, along with the aforementioned event selections, the event efficiency is evaluated per science run by tracking the survival ratio of simulated events randomly inserted into real data assuming a spatially and temporally uniform signal distribution \cite{axidence}. The event efficiency is 77\% (73\%) for 2-hit S1, and 82\% (78\%) for 3-hit S1 in SR0 (SR1) within the ROI.

\itsec{Backgrounds} This analysis considers the background described in Refs. \cite{XENON:2020gfr,XENON:2023cxc,XENON:2024xgd}. The expected rate and systematic uncertainties, both in the ROI and in the signal-like region, which is defined as the region containing 50\% of the signal for bins ranked by signal-to-background ratio, are summarized in Table \ref{tab:table1}.

The solar \beight ~\cevns interaction, whose recoil spectrum is nearly indistinguishable from a 5.5 \gevcsq ~DM-nucleon scattering at cross-section \SI{4.4e-45}{\cm^{2}}, is one of the dominant backgrounds of this search. Following the recommended convention \cite{Baxter:2021pqo}, the predicted solar \beight ~\cevns rate after all selections is $\SI[inter-unit-product=\;]{3.6 \pm 0.9}{\eventspertonyear}$ in SR0 and \SI{3.3 \pm 0.8}{\eventspertonyear} in SR1, based on the standard solar model \beight ~neutrino spectrum \cite{Bahcall:1996qv} and the measured flux of \SI{5.25 \pm 0.20e6}{\per\cm\squared\per\second} from SNO \cite{SNO:2011hxd}, using the nominal NR yield model.

The accidental coincidence (AC) background arises from the accidental pairing of isolated S1 and isolated S2 peaks, where `isolated' refers to peaks that lack a physically corresponding S1 or S2 \cite{XENON:2024ijk,axidence}. The quality and reconstruction selections outlined in the signal section reduce the isolated S1 and S2 peak rates to 2.3 (2.2) Hz for S1 and 18 (26) mHz for S2 in SR0 (SR1). Utilizing a synthesized data-driven AC dataset \cite{XENON:2024xgd,XENON:2020gfr}, two boosted decision tree (BDT) classifiers, S1 BDT and S2 BDT, are trained on an equivalent 6 \gevcsq ~DM signal, each targeting AC of different origins. The S1 BDT distinguishes isolated S1 signals originating from random PMT hit clustering by incorporating S1 hit counts, PMT channel distribution, double photo-electron emission \cite{Faham:2015kqa}, and S1 pulse shape \cite{Akerib:2021pfd} into the S1 BDT score. The S2 BDT evaluates whether the S2 pulse shape of the candidate event is compatible with the correlation between electron diffusion and the drift time \cite{sorensen_anisotropic_2011}, converting this assessment into an S2 BDT score. These waveform-feature-based BDT scores are used as an inference dimension to enhance signal-to-background discrimination power, as shown in Fig. \ref{fig:analysis_dimension}. 
An AC sideband, primarily consisting of events rejected by the S2 BDT selection, was unblinded before the ROI unblinding to examine the validity of the AC modeling. Subsequently, the S2 area threshold of the ROI was raised from 100 PE to 120 PE due to an observed tension between the observation and the AC background model. The discrepancy in the AC rate between the unblinding result of the AC sideband and the prediction is then less than 5\%, and the statistical uncertainty of the AC sideband data is conservatively chosen to be the systematic uncertainty of the AC background. The details of sideband validation can be found in Ref. \cite{XENON:2024ijk}. 
The AC background prediction is \SI{6.4\pm0.6}{\eventspertonyear} in SR0 and \SI{7.6\pm0.4}{\eventspertonyear} in SR1.

\begin{figure*}[htbp]
\centering
\renewcommand{\arraystretch}{0} % This reduces the default spacing
\begin{tabular}{ >{\centering\arraybackslash}m{8.8cm}  >{\centering\arraybackslash}m{8.63cm}  }
    \includegraphics[width=8.8cm]{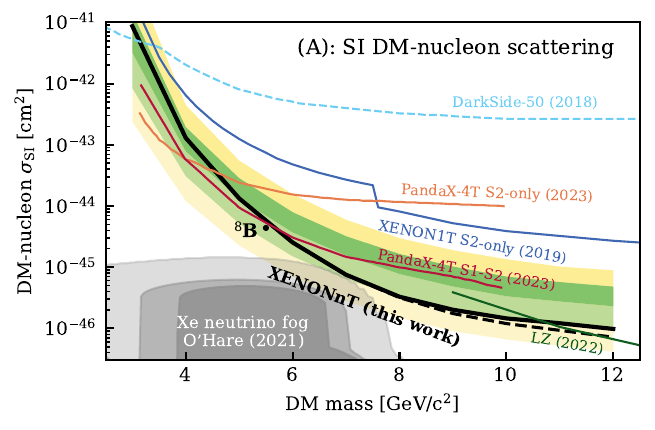} & 
    \raisebox{0.03cm}{\includegraphics[width=8.61cm]{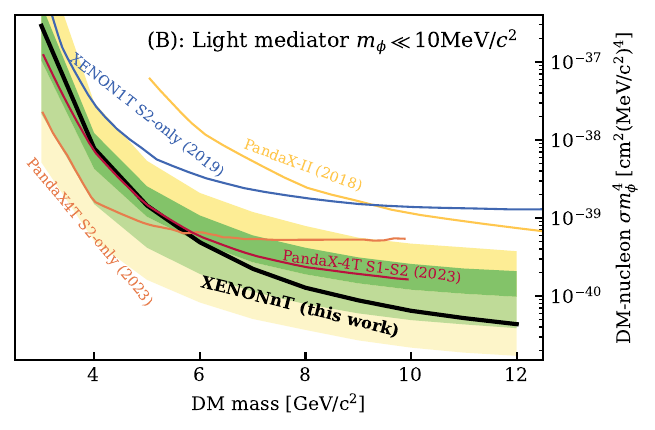}} \\
    \noalign{\vskip -0.2cm} % Adjust the negative space here
    \includegraphics[width=8.7cm]{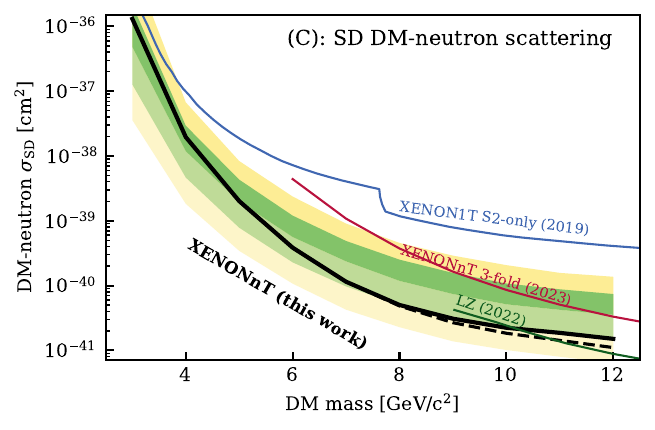} & 
    \raisebox{0.03cm}{\includegraphics[width=8.63cm]{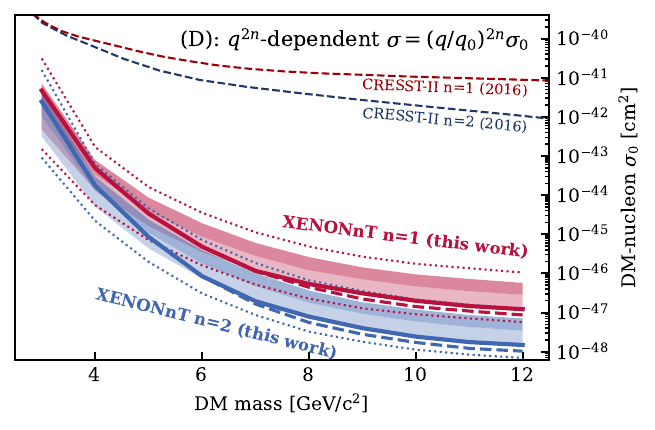}} \\
\end{tabular}

\caption{
DM cross section upper limit at 90\% confidence level with the 1$\sigma$ (green) and 2$\sigma$ (yellow) sensitivity band. The results with and without sensitivity -1$\sigma$ PCL are shown by solid and dashed lines, respectively. Previous results are shown for comparison \cite{XENON:2020gfr, XENON:2019gfn,LZ:2022lsv,PandaX:2022xqx,PandaX:2022aac,Angloher:2016jsl,DarkSide-50:2022qzh,PandaX:2023xgl}, where limits from other liquid-xenon-based experiments are power-constrained to sensitivity -1$\sigma$.
(A) shows the constraint on SI DM-nucleon interaction. The neutrino fog \cite{OHare:2021utq} is defined as $n = \max_{\sigma \in [\sigma, \infty]} \left[ -\left( \frac{d \ln \sigma}{d \ln MT} \right)^{-1} \right]$, in which $n=2,2.5,3$ from lighter to darker gray. The \beight ~\cevns ~equivalent DM of 5.5 \gevcsq~with cross section \SI{4.4e-45}{cm^{2}} is shown. 
(B) shows the constraints on SI-LM, where the differential rate is proportional to $\sigma m_\phi{ }^4$ assuming $m_\phi \ll 10 ~\mathrm{MeV}/c^2$  \cite{nobile_direct_2015,XENON:2019gfn}.
(C) shows the constraint on SD DM-neutron interaction assuming the median of nuclear form factors from Ref. \cite{klos_large-scale_2013}. %The limit is susceptible to the current 10$\%$ uncertainty of the nuclear form factors in Ref. \cite{klos_large-scale_2013}.
(D) shows the constraint on MDDM with momentum dependence $n=1,2$ (red, blue), where $\sigma_0$ is the SI cross section. The 1 and 2$\sigma$ sensitivity bands are depicted using shadow and dots, respectively.
}
\label{fig:DM_limits}
\end{figure*}

The surface background originates from $^{222}\mathrm{Rn}$ progeny plate-out on the PTFE panels, such as $\beta$ decays from $^{210}\mathrm{Pb}$ ($\mathrm{\bar{E}_R} \approx 6.1~\mathrm{keV_{ER}}$). This can result in a higher background rate in the ROI due to the presence of S2 charge-insensitive volumes and the worse position reconstruction resolution of small S2 signals ($\Delta R \approx \SI{2}{\cm} $) near the detector wall. The surface background is modeled in a data-driven manner using non-blinded events. The FV is selected to mitigate the surface background contribution, yielding a negligible rate of less than \SI{0.10}{\eventspertonyear} at 90\% CL. The prediction is validated by a sideband unblinding of events within $\SI[inter-unit-product=\;]{61.4}{\cm}<R<\SI[inter-unit-product=\;]{63.0}{\cm}$. Due to the negligible rate, the surface background is not included in the statistical model.

The radiogenic neutron background originates from nuclear fission and the $(\alpha,\mathrm{n})$ process in the uranium and thorium decay chains. It is modeled using full-chain simulations as described in Ref. \cite{XENON:2024xgd}. The levels of the radioactive contamination, as determined from the radioassay \cite{XENON:2023cxc}, are further adjusted based on the comparison between observed and simulated high energy $\gamma$ ray activity. The prediction of the neutron background is \SI{0.11\pm0.06}{\eventspertonyear} in SR0 and \SI{0.14\pm0.08}{\eventspertonyear} in SR1, considering an effective neutron veto tagging efficiency of 49\% (SR0) and 52\% (SR1).

The electronic recoil background originates from internal radioactive contaminations, including the $\beta$ decays of $^{214}\mathrm{Pb}$, $^{85}\mathrm{Kr}$, material $\gamma$ rays, and the neutrino-electron weak interaction of solar $pp$ neutrinos \cite{XENON:2022ltv}. The ER rates per SR are determined by fitting the already unblinded events in the [1, 140] keV range in SR0 \cite{XENON:2022ltv} and the non-blinded events in the [10, 140] range keV in SR1. Due to the uncertainty of the shape of the ER background energy spectrum, a systematic uncertainty of 100\% is assigned to the ER background, under the assumption of a flat ER energy spectrum in the ROI and utilizing the combined fit yield model from SR0 $^{220}\mathrm{Rn}$ and $^{37}\mathrm{Ar}$ calibration data \cite{XENON:2024xgd}. The conservative uncertainty has minimal impact on our results due to the low rate of ER background. The prediction of ER background is \SI{0.11\pm0.11}{\eventspertonyear} in SR0 and \SI{0.24\pm0.24}{\eventspertonyear} in SR1, and the difference between SRs is due to an increased concentration of $^{85}\mathrm{Kr}$ at the start of SR1 which was subsequently removed by cryogenic distillation.

\itsec{Inference} An extended binned likelihood function is constructed for the SR0 and SR1 data using four analysis dimensions: corrected S2 (cS2) \cite{XENON:2024qgt}, S1 BDT score, S2 BDT score, and $\text{S2}_{\text{pre}}/\Delta t_{\text{pre}}$. The bin edges are defined by binning the AC background in each dimension with equal probability, as shown in Fig. \ref{fig:analysis_dimension}, resulting in a total of $3^4=81$ bins in the 4D histogram. The likelihood function is expressed as:
\begin{equation}\label{eqn:likelihood}
\mathcal{L}(\sigma, \boldsymbol{\theta}) = \mathcal{L}_\mathrm{SR0}(\sigma, \boldsymbol{\theta}) \cdot \mathcal{L}_\mathrm{SR1}(\sigma, \boldsymbol{\theta}) \cdot \prod_m \mathcal{L}^{\text{anc}}_m(\theta_m),
\end{equation}
where $\sigma$ represents the DM-nucleon interaction cross section, and $\boldsymbol{\theta}$ are the Gaussian ancillary likelihood-constrained nuisance parameters. The index $m$ iterates through the nuisance parameters, including the expected rate of each background with their uncertainties shown in Table. \ref{tab:table1}, the signal efficiency with 25\% uncertainty, and the NR yield model shape parameters $(t_{\mathrm{Ly}}, t_{\mathrm{Qy}})$. 

Prior to unblinding, the sensitivity band was determined by computing the profile-likelihood-ratio test statistics between the signal-plus-background hypothesis and the background-only hypothesis using toy Monte-Carlo datasets \cite{Feldman:1997qc,Baxter:2021pqo,alea}. We adopt a power-constrained limit (PCL) \cite{Cowan:2011an} with a power threshold of $-1\sigma$ to report the final limit. The power threshold is changed from our previous 50\% threshold \cite{XENON:2023cxc} to be in line with the limits reported by other experiments. A future publication will discuss the difference between PCL thresholds and further motivate this choice.

\itsec{Results} After unblinding, SR0 and SR1 reveal 9 and 28 events, respectively. The discovery p-value for the DM of SI interaction indicates no significant excess, with a minimum p-value of $0.18$ at 3 \gevcsq. Their distributions in the search space are shown in Fig. \ref{fig:analysis_dimension}. The background-only and signal-plus-background best-fits for 4, 6, 10 \gevcsq ~DMs are presented in Table. \ref{tab:table1}. Two goodness-of-fit (GOF) tests, with a reporting p-value threshold of 0.025 determined prior to unblinding, are performed on the spatial $X$-$Y$ dimension. The resulting p-values, 0.018 and 0.037, suggest potential spatial clustering of events. However, the positions are not used in the inference framework defined prior to unblinding. Post-unblinding checks including inspection of the distribution in data selection space as well as the event waveforms show no indication of mismodeling.

The results for SI, SD, MDDM, and SI-LM, along with previous limits \cite{XENON:2020gfr,XENON:2019gfn,LZ:2022lsv,PandaX:2022xqx,PandaX:2022aac,Angloher:2016jsl,DarkSide-50:2022qzh,PandaX:2023xgl}, are presented in Fig. \ref{fig:DM_limits} for comparison. Our results tighten the constraints across the parameter spaces for all listed DM models, with the SI limit at the benchmark 6 \gevcsq ~DM mass reaching \SI{2.5e-45}{cm^{2}}. An upward fluctuation is observed below 5 \gevcsq ~and a downward fluctuation above this mass.  The SD DM-neutron limit is susceptible to the current $\sim10\%$ uncertainty of the nuclear form factors in Ref. \cite{klos_large-scale_2013}. As the uncertainty predominately changes the expected recoil rate, the results presented here can be re-scaled. Additionally, we exclude all metallic components of mirror DM proposed by Ref. \cite{Clarke:2016eac,Foot:2014mia}, specifically ruling out mirror oxygen with a 90\% upper limit on the rate parameter $\epsilon \sqrt{\xi_{O^{\prime}}}$ of $1.3 \times 10^{-12}$, with the theoretically favored region $10^{-11} \lesssim \epsilon \sqrt{\xi_{O^{\prime}}} \lesssim 4 \times 10^{-10}$.

XENONnT achieves the world-leading sensitivity for light DM candidates within the \beight ~\cevns neutrino fog, as illustrated by Fig. \ref{fig:DM_limits} (A). The background-only model yields a best-fit of $11.4^{+2.0}_{-3.6}$ \beight ~\cevns events. In the 6 \gevcsq ~model, \beight ~\cevns emerges as the most significant background, accounting for 60\% of the total background in the signal-like region, marking the first step into the neutrino fog. We have prepared a dedicated data release to allow further interpretation of this work with different DM models and yield models in Ref. \cite{lightwimp_data_release}.

\itsec{Acknowledgements} 
We gratefully acknowledge support from the National Science Foundation, Swiss National Science Foundation, German Ministry for Education and Research, Max Planck Gesellschaft, Deutsche Forschungsgemeinschaft, Helmholtz Association, Dutch Research Council (NWO), Fundacao para a Ciencia e Tecnologia, Weizmann Institute of Science, Binational Science Foundation, Région des Pays de la Loire, Knut and Alice Wallenberg Foundation, Kavli Foundation, JSPS Kakenhi, JST FOREST Program, and ERAN in Japan, Tsinghua University Initiative Scientific Research Program, DIM-ACAV+ Région Ile-de-France, and Istituto Nazionale di Fisica Nucleare. This project has received funding/support from the European Union’s Horizon 2020 research and innovation program under the Marie Skłodowska-Curie grant agreement No 860881-HIDDeN.

We gratefully acknowledge support for providing computing and data-processing resources of the Open Science Pool and the European Grid Initiative, at the following computing centers: the CNRS/IN2P3 (Lyon - France), the Dutch national e-infrastructure with the support of SURF Cooperative, the Nikhef Data-Processing Facility (Amsterdam - Netherlands), the INFN-CNAF (Bologna - Italy), the San Diego Supercomputer Center (San Diego - USA) and the Enrico Fermi Institute (Chicago - USA). We acknowledge the support of the Research Computing Center (RCC) at The University of Chicago for providing computing resources for data analysis.

We thank the INFN Laboratori Nazionali del Gran Sasso for hosting and supporting the XENON project.

\bibliography{main}

\begin{thebibliography}{61}%
\makeatletter
\providecommand \@ifxundefined [1]{%
 \@ifx{#1\undefined}
}%
\providecommand \@ifnum [1]{%
 \ifnum #1\expandafter \@firstoftwo
 \else \expandafter \@secondoftwo
 \fi
}%
\providecommand \@ifx [1]{%
 \ifx #1\expandafter \@firstoftwo
 \else \expandafter \@secondoftwo
 \fi
}%
\providecommand \natexlab [1]{#1}%
\providecommand \enquote  [1]{``#1''}%
\providecommand \bibnamefont  [1]{#1}%
\providecommand \bibfnamefont [1]{#1}%
\providecommand \citenamefont [1]{#1}%
\providecommand \href@noop [0]{\@secondoftwo}%
\providecommand \href [0]{\begingroup \@sanitize@url \@href}%
\providecommand \@href[1]{\@@startlink{#1}\@@href}%
\providecommand \@@href[1]{\endgroup#1\@@endlink}%
\providecommand \@sanitize@url [0]{\catcode `\\12\catcode `\$12\catcode `\&12\catcode `\#12\catcode `\^12\catcode `\_12\catcode `\%12\relax}%
\providecommand \@@startlink[1]{}%
\providecommand \@@endlink[0]{}%
\providecommand \url  [0]{\begingroup\@sanitize@url \@url }%
\providecommand \@url [1]{\endgroup\@href {#1}{\urlprefix }}%
\providecommand \urlprefix  [0]{URL }%
\providecommand \Eprint [0]{\href }%
\providecommand \doibase [0]{https://doi.org/}%
\providecommand \selectlanguage [0]{\@gobble}%
\providecommand \bibinfo  [0]{\@secondoftwo}%
\providecommand \bibfield  [0]{\@secondoftwo}%
\providecommand \translation [1]{[#1]}%
\providecommand \BibitemOpen [0]{}%
\providecommand \bibitemStop [0]{}%
\providecommand \bibitemNoStop [0]{.\EOS\space}%
\providecommand \EOS [0]{\spacefactor3000\relax}%
\providecommand \BibitemShut  [1]{\csname bibitem#1\endcsname}%
\let\auto@bib@innerbib\@empty
%</preamble>
\bibitem [{\citenamefont {Aprile}\ \emph {et~al.}(2024{\natexlab{a}})\citenamefont {Aprile} \emph {et~al.}}]{XENON:2024wpa}%
  \BibitemOpen
  \bibfield  {author} {\bibinfo {author} {\bibfnamefont {E.}~\bibnamefont {Aprile}} \emph {et~al.} (\bibinfo {collaboration} {XENON}),\ }\bibfield  {title} {\bibinfo {title} {{The XENONnT dark matter experiment}},\ }\href {https://doi.org/10.1140/epjc/s10052-024-12982-5} {\bibfield  {journal} {\bibinfo  {journal} {Eur. Phys. J. C}\ }\textbf {\bibinfo {volume} {84}},\ \bibinfo {pages} {784} (\bibinfo {year} {2024}{\natexlab{a}})},\ \Eprint {https://arxiv.org/abs/2402.10446} {arXiv:2402.10446 [physics.ins-det]} \BibitemShut {NoStop}%
\bibitem [{\citenamefont {Roszkowski}\ \emph {et~al.}(2018)\citenamefont {Roszkowski}, \citenamefont {Sessolo},\ and\ \citenamefont {Trojanowski}}]{Roszkowski:2017nbc}%
  \BibitemOpen
  \bibfield  {author} {\bibinfo {author} {\bibfnamefont {L.}~\bibnamefont {Roszkowski}}, \bibinfo {author} {\bibfnamefont {E.~M.}\ \bibnamefont {Sessolo}},\ and\ \bibinfo {author} {\bibfnamefont {S.}~\bibnamefont {Trojanowski}},\ }\bibfield  {title} {\bibinfo {title} {{WIMP dark matter candidates and searches\textemdash{}current status and future prospects}},\ }\href {https://doi.org/10.1088/1361-6633/aab913} {\bibfield  {journal} {\bibinfo  {journal} {Rept. Prog. Phys.}\ }\textbf {\bibinfo {volume} {81}},\ \bibinfo {pages} {066201} (\bibinfo {year} {2018})},\ \Eprint {https://arxiv.org/abs/1707.06277} {arXiv:1707.06277 [hep-ph]} \BibitemShut {NoStop}%
\bibitem [{\citenamefont {Jungman}\ \emph {et~al.}(1996)\citenamefont {Jungman}, \citenamefont {Kamionkowski},\ and\ \citenamefont {Griest}}]{Jungman:1995df}%
  \BibitemOpen
  \bibfield  {author} {\bibinfo {author} {\bibfnamefont {G.}~\bibnamefont {Jungman}}, \bibinfo {author} {\bibfnamefont {M.}~\bibnamefont {Kamionkowski}},\ and\ \bibinfo {author} {\bibfnamefont {K.}~\bibnamefont {Griest}},\ }\bibfield  {title} {\bibinfo {title} {{Supersymmetric dark matter}},\ }\href {https://doi.org/10.1016/0370-1573(95)00058-5} {\bibfield  {journal} {\bibinfo  {journal} {Phys. Rept.}\ }\textbf {\bibinfo {volume} {267}},\ \bibinfo {pages} {195} (\bibinfo {year} {1996})},\ \Eprint {https://arxiv.org/abs/hep-ph/9506380} {arXiv:hep-ph/9506380} \BibitemShut {NoStop}%
\bibitem [{\citenamefont {Bertone}\ and\ \citenamefont {Hooper}(2018)}]{Bertone:2016nfn}%
  \BibitemOpen
  \bibfield  {author} {\bibinfo {author} {\bibfnamefont {G.}~\bibnamefont {Bertone}}\ and\ \bibinfo {author} {\bibfnamefont {D.}~\bibnamefont {Hooper}},\ }\bibfield  {title} {\bibinfo {title} {{History of dark matter}},\ }\href {https://doi.org/10.1103/RevModPhys.90.045002} {\bibfield  {journal} {\bibinfo  {journal} {Rev. Mod. Phys.}\ }\textbf {\bibinfo {volume} {90}},\ \bibinfo {pages} {045002} (\bibinfo {year} {2018})},\ \Eprint {https://arxiv.org/abs/1605.04909} {arXiv:1605.04909 [astro-ph.CO]} \BibitemShut {NoStop}%
\bibitem [{\citenamefont {Aprile}\ \emph {et~al.}(2023)\citenamefont {Aprile} \emph {et~al.}}]{XENON:2023cxc}%
  \BibitemOpen
  \bibfield  {author} {\bibinfo {author} {\bibfnamefont {E.}~\bibnamefont {Aprile}} \emph {et~al.} (\bibinfo {collaboration} {XENON}),\ }\bibfield  {title} {\bibinfo {title} {{First Dark Matter Search with Nuclear Recoils from the XENONnT Experiment}},\ }\href {https://doi.org/10.1103/PhysRevLett.131.041003} {\bibfield  {journal} {\bibinfo  {journal} {Phys. Rev. Lett.}\ }\textbf {\bibinfo {volume} {131}},\ \bibinfo {pages} {041003} (\bibinfo {year} {2023})},\ \Eprint {https://arxiv.org/abs/2303.14729} {arXiv:2303.14729 [hep-ex]} \BibitemShut {NoStop}%
\bibitem [{\citenamefont {O'Hare}(2021)}]{OHare:2021utq}%
  \BibitemOpen
  \bibfield  {author} {\bibinfo {author} {\bibfnamefont {C.~A.~J.}\ \bibnamefont {O'Hare}},\ }\bibfield  {title} {\bibinfo {title} {{New Definition of the Neutrino Floor for Direct Dark Matter Searches}},\ }\href {https://doi.org/10.1103/PhysRevLett.127.251802} {\bibfield  {journal} {\bibinfo  {journal} {Phys. Rev. Lett.}\ }\textbf {\bibinfo {volume} {127}},\ \bibinfo {pages} {251802} (\bibinfo {year} {2021})},\ \Eprint {https://arxiv.org/abs/2109.03116} {arXiv:2109.03116 [hep-ph]} \BibitemShut {NoStop}%
\bibitem [{\citenamefont {Billard}\ \emph {et~al.}(2014)\citenamefont {Billard}, \citenamefont {Strigari},\ and\ \citenamefont {Figueroa-Feliciano}}]{Billard:2013qya}%
  \BibitemOpen
  \bibfield  {author} {\bibinfo {author} {\bibfnamefont {J.}~\bibnamefont {Billard}}, \bibinfo {author} {\bibfnamefont {L.}~\bibnamefont {Strigari}},\ and\ \bibinfo {author} {\bibfnamefont {E.}~\bibnamefont {Figueroa-Feliciano}},\ }\bibfield  {title} {\bibinfo {title} {{Implication of neutrino backgrounds on the reach of next generation dark matter direct detection experiments}},\ }\href {https://doi.org/10.1103/PhysRevD.89.023524} {\bibfield  {journal} {\bibinfo  {journal} {Phys. Rev. D}\ }\textbf {\bibinfo {volume} {89}},\ \bibinfo {pages} {023524} (\bibinfo {year} {2014})},\ \Eprint {https://arxiv.org/abs/1307.5458} {arXiv:1307.5458 [hep-ph]} \BibitemShut {NoStop}%
\bibitem [{\citenamefont {Ruppin}\ \emph {et~al.}(2014)\citenamefont {Ruppin}, \citenamefont {Billard}, \citenamefont {Figueroa-Feliciano},\ and\ \citenamefont {Strigari}}]{Ruppin:2014bra}%
  \BibitemOpen
  \bibfield  {author} {\bibinfo {author} {\bibfnamefont {F.}~\bibnamefont {Ruppin}}, \bibinfo {author} {\bibfnamefont {J.}~\bibnamefont {Billard}}, \bibinfo {author} {\bibfnamefont {E.}~\bibnamefont {Figueroa-Feliciano}},\ and\ \bibinfo {author} {\bibfnamefont {L.}~\bibnamefont {Strigari}},\ }\bibfield  {title} {\bibinfo {title} {{Complementarity of dark matter detectors in light of the neutrino background}},\ }\href {https://doi.org/10.1103/PhysRevD.90.083510} {\bibfield  {journal} {\bibinfo  {journal} {Phys. Rev. D}\ }\textbf {\bibinfo {volume} {90}},\ \bibinfo {pages} {083510} (\bibinfo {year} {2014})},\ \Eprint {https://arxiv.org/abs/1408.3581} {arXiv:1408.3581 [hep-ph]} \BibitemShut {NoStop}%
\bibitem [{\citenamefont {Akerib}\ \emph {et~al.}(2022)\citenamefont {Akerib} \emph {et~al.}}]{Akerib:2022ort}%
  \BibitemOpen
  \bibfield  {author} {\bibinfo {author} {\bibfnamefont {D.~S.}\ \bibnamefont {Akerib}} \emph {et~al.},\ }\bibfield  {title} {\bibinfo {title} {{Snowmass2021 Cosmic Frontier Dark Matter Direct Detection to the Neutrino Fog}},\ }in\ \href@noop {} {\emph {\bibinfo {booktitle} {{Snowmass 2021}}}}\ (\bibinfo {year} {2022})\ \Eprint {https://arxiv.org/abs/2203.08084} {arXiv:2203.08084 [hep-ex]} \BibitemShut {NoStop}%
\bibitem [{\citenamefont {Billard}\ \emph {et~al.}(2022)\citenamefont {Billard} \emph {et~al.}}]{Billard:2021uyg}%
  \BibitemOpen
  \bibfield  {author} {\bibinfo {author} {\bibfnamefont {J.}~\bibnamefont {Billard}} \emph {et~al.},\ }\bibfield  {title} {\bibinfo {title} {{Direct detection of dark matter\textemdash{}APPEC committee report*}},\ }\href {https://doi.org/10.1088/1361-6633/ac5754} {\bibfield  {journal} {\bibinfo  {journal} {Rept. Prog. Phys.}\ }\textbf {\bibinfo {volume} {85}},\ \bibinfo {pages} {056201} (\bibinfo {year} {2022})},\ \Eprint {https://arxiv.org/abs/2104.07634} {arXiv:2104.07634 [hep-ex]} \BibitemShut {NoStop}%
\bibitem [{\citenamefont {Aprile}\ \emph {et~al.}(2024{\natexlab{b}})\citenamefont {Aprile} \emph {et~al.}}]{XENON:2024ijk}%
  \BibitemOpen
  \bibfield  {author} {\bibinfo {author} {\bibfnamefont {E.}~\bibnamefont {Aprile}} \emph {et~al.} (\bibinfo {collaboration} {XENON}),\ }\bibfield  {title} {\bibinfo {title} {{First Indication of Solar $^8B$~Neutrinos via Coherent Elastic Neutrino-Nucleus Scattering with XENONnT}},\ }\href {https://doi.org/10.1103/PhysRevLett.133.191002} {\bibfield  {journal} {\bibinfo  {journal} {Phys. Rev. Lett.}\ }\textbf {\bibinfo {volume} {133}},\ \bibinfo {pages} {191002} (\bibinfo {year} {2024}{\natexlab{b}})},\ \Eprint {https://arxiv.org/abs/2408.02877} {arXiv:2408.02877 [nucl-ex]} \BibitemShut {NoStop}%
\bibitem [{\citenamefont {Aprile}\ \emph {et~al.}(2014)\citenamefont {Aprile} \emph {et~al.}}]{XENON1T:2014eqx}%
  \BibitemOpen
  \bibfield  {author} {\bibinfo {author} {\bibfnamefont {E.}~\bibnamefont {Aprile}} \emph {et~al.} (\bibinfo {collaboration} {XENON1T}),\ }\bibfield  {title} {\bibinfo {title} {{Conceptual design and simulation of a water Cherenkov muon veto for the XENON1T experiment}},\ }\href {https://doi.org/10.1088/1748-0221/9/11/P11006} {\bibfield  {journal} {\bibinfo  {journal} {JINST}\ }\textbf {\bibinfo {volume} {9}},\ \bibinfo {pages} {P11006}},\ \Eprint {https://arxiv.org/abs/1406.2374} {arXiv:1406.2374 [astro-ph.IM]} \BibitemShut {NoStop}%
\bibitem [{\citenamefont {Murra}\ \emph {et~al.}(2022)\citenamefont {Murra}, \citenamefont {Schulte}, \citenamefont {Huhmann},\ and\ \citenamefont {Weinheimer}}]{Murra:2022mlr}%
  \BibitemOpen
  \bibfield  {author} {\bibinfo {author} {\bibfnamefont {M.}~\bibnamefont {Murra}}, \bibinfo {author} {\bibfnamefont {D.}~\bibnamefont {Schulte}}, \bibinfo {author} {\bibfnamefont {C.}~\bibnamefont {Huhmann}},\ and\ \bibinfo {author} {\bibfnamefont {C.}~\bibnamefont {Weinheimer}},\ }\bibfield  {title} {\bibinfo {title} {{Design, construction and commissioning of a high-flow radon removal system for XENONnT}},\ }\href {https://doi.org/10.1140/epjc/s10052-022-11001-9} {\bibfield  {journal} {\bibinfo  {journal} {Eur. Phys. J. C}\ }\textbf {\bibinfo {volume} {82}},\ \bibinfo {pages} {1104} (\bibinfo {year} {2022})},\ \Eprint {https://arxiv.org/abs/2205.11492} {arXiv:2205.11492 [physics.ins-det]} \BibitemShut {NoStop}%
\bibitem [{\citenamefont {Aprile}\ \emph {et~al.}(2017)\citenamefont {Aprile} \emph {et~al.}}]{XENON:2016bmq}%
  \BibitemOpen
  \bibfield  {author} {\bibinfo {author} {\bibfnamefont {E.}~\bibnamefont {Aprile}} \emph {et~al.} (\bibinfo {collaboration} {XENON}),\ }\bibfield  {title} {\bibinfo {title} {{Removing krypton from xenon by cryogenic distillation to the ppq level}},\ }\href {https://doi.org/10.1140/epjc/s10052-017-4757-1} {\bibfield  {journal} {\bibinfo  {journal} {Eur. Phys. J. C}\ }\textbf {\bibinfo {volume} {77}},\ \bibinfo {pages} {275} (\bibinfo {year} {2017})},\ \Eprint {https://arxiv.org/abs/1612.04284} {arXiv:1612.04284 [physics.ins-det]} \BibitemShut {NoStop}%
\bibitem [{\citenamefont {Aalbers}\ \emph {et~al.}(2024)\citenamefont {Aalbers} \emph {et~al.}}]{strax}%
  \BibitemOpen
  \bibfield  {author} {\bibinfo {author} {\bibfnamefont {J.}~\bibnamefont {Aalbers}} \emph {et~al.},\ }\href {https://doi.org/10.5281/zenodo.11355772} {\bibinfo {title} {{AxFoundation}/strax: Stream analysis for xenon tpcs}} (\bibinfo {year} {2024})\BibitemShut {NoStop}%
\bibitem [{\citenamefont {{XENON Collaboration}}(2024{\natexlab{a}})}]{straxen}%
  \BibitemOpen
  \bibfield  {author} {\bibinfo {author} {\bibnamefont {{XENON Collaboration}}},\ }\href {https://doi.org/10.5281/zenodo.12608732} {\bibinfo {title} {{XENONnT}/straxen: Streaming analysis for xenon}} (\bibinfo {year} {2024}{\natexlab{a}})\BibitemShut {NoStop}%
\bibitem [{\citenamefont {Tom\'as}\ \emph {et~al.}(2018)\citenamefont {Tom\'as}, \citenamefont {Ara\'ujo}, \citenamefont {Bailey}, \citenamefont {Bayer}, \citenamefont {Chen}, \citenamefont {L\'opez~Paredes},\ and\ \citenamefont {Sumner}}]{Tomas:2018pny}%
  \BibitemOpen
  \bibfield  {author} {\bibinfo {author} {\bibfnamefont {A.}~\bibnamefont {Tom\'as}}, \bibinfo {author} {\bibfnamefont {H.~M.}\ \bibnamefont {Ara\'ujo}}, \bibinfo {author} {\bibfnamefont {A.~J.}\ \bibnamefont {Bailey}}, \bibinfo {author} {\bibfnamefont {A.}~\bibnamefont {Bayer}}, \bibinfo {author} {\bibfnamefont {E.}~\bibnamefont {Chen}}, \bibinfo {author} {\bibfnamefont {B.}~\bibnamefont {L\'opez~Paredes}},\ and\ \bibinfo {author} {\bibfnamefont {T.~J.}\ \bibnamefont {Sumner}},\ }\bibfield  {title} {\bibinfo {title} {{Study and mitigation of spurious electron emission from cathodic wires in noble liquid time projection chambers}},\ }\href {https://doi.org/10.1016/j.astropartphys.2018.07.001} {\bibfield  {journal} {\bibinfo  {journal} {Astropart. Phys.}\ }\textbf {\bibinfo {volume} {103}},\ \bibinfo {pages} {49} (\bibinfo {year} {2018})},\ \Eprint {https://arxiv.org/abs/1801.07231} {arXiv:1801.07231 [physics.ins-det]} \BibitemShut {NoStop}%
\bibitem [{\citenamefont {Akerib}\ \emph {et~al.}(2020)\citenamefont {Akerib} \emph {et~al.}}]{LUX:2020vbj}%
  \BibitemOpen
  \bibfield  {author} {\bibinfo {author} {\bibfnamefont {D.~S.}\ \bibnamefont {Akerib}} \emph {et~al.} (\bibinfo {collaboration} {LUX}),\ }\bibfield  {title} {\bibinfo {title} {{Investigation of background electron emission in the LUX detector}},\ }\href {https://doi.org/10.1103/PhysRevD.102.092004} {\bibfield  {journal} {\bibinfo  {journal} {Phys. Rev. D}\ }\textbf {\bibinfo {volume} {102}},\ \bibinfo {pages} {092004} (\bibinfo {year} {2020})},\ \Eprint {https://arxiv.org/abs/2004.07791} {arXiv:2004.07791 [physics.ins-det]} \BibitemShut {NoStop}%
\bibitem [{\citenamefont {Petraki}\ and\ \citenamefont {Volkas}(2013)}]{Petraki:2013wwa}%
  \BibitemOpen
  \bibfield  {author} {\bibinfo {author} {\bibfnamefont {K.}~\bibnamefont {Petraki}}\ and\ \bibinfo {author} {\bibfnamefont {R.~R.}\ \bibnamefont {Volkas}},\ }\bibfield  {title} {\bibinfo {title} {{Review of asymmetric dark matter}},\ }\href {https://doi.org/10.1142/S0217751X13300287} {\bibfield  {journal} {\bibinfo  {journal} {Int. J. Mod. Phys. A}\ }\textbf {\bibinfo {volume} {28}},\ \bibinfo {pages} {1330028} (\bibinfo {year} {2013})},\ \Eprint {https://arxiv.org/abs/1305.4939} {arXiv:1305.4939 [hep-ph]} \BibitemShut {NoStop}%
\bibitem [{\citenamefont {Kaplan}\ \emph {et~al.}(2009)\citenamefont {Kaplan}, \citenamefont {Luty},\ and\ \citenamefont {Zurek}}]{Kaplan:2009ag}%
  \BibitemOpen
  \bibfield  {author} {\bibinfo {author} {\bibfnamefont {D.~E.}\ \bibnamefont {Kaplan}}, \bibinfo {author} {\bibfnamefont {M.~A.}\ \bibnamefont {Luty}},\ and\ \bibinfo {author} {\bibfnamefont {K.~M.}\ \bibnamefont {Zurek}},\ }\bibfield  {title} {\bibinfo {title} {{Asymmetric Dark Matter}},\ }\href {https://doi.org/10.1103/PhysRevD.79.115016} {\bibfield  {journal} {\bibinfo  {journal} {Phys. Rev. D}\ }\textbf {\bibinfo {volume} {79}},\ \bibinfo {pages} {115016} (\bibinfo {year} {2009})},\ \Eprint {https://arxiv.org/abs/0901.4117} {arXiv:0901.4117 [hep-ph]} \BibitemShut {NoStop}%
\bibitem [{\citenamefont {Spergel}\ and\ \citenamefont {Steinhardt}(2000)}]{Spergel:1999mh}%
  \BibitemOpen
  \bibfield  {author} {\bibinfo {author} {\bibfnamefont {D.~N.}\ \bibnamefont {Spergel}}\ and\ \bibinfo {author} {\bibfnamefont {P.~J.}\ \bibnamefont {Steinhardt}},\ }\bibfield  {title} {\bibinfo {title} {{Observational evidence for selfinteracting cold dark matter}},\ }\href {https://doi.org/10.1103/PhysRevLett.84.3760} {\bibfield  {journal} {\bibinfo  {journal} {Phys. Rev. Lett.}\ }\textbf {\bibinfo {volume} {84}},\ \bibinfo {pages} {3760} (\bibinfo {year} {2000})},\ \Eprint {https://arxiv.org/abs/astro-ph/9909386} {arXiv:astro-ph/9909386} \BibitemShut {NoStop}%
\bibitem [{\citenamefont {Tulin}\ and\ \citenamefont {Yu}(2018)}]{Tulin:2017ara}%
  \BibitemOpen
  \bibfield  {author} {\bibinfo {author} {\bibfnamefont {S.}~\bibnamefont {Tulin}}\ and\ \bibinfo {author} {\bibfnamefont {H.-B.}\ \bibnamefont {Yu}},\ }\bibfield  {title} {\bibinfo {title} {{Dark Matter Self-interactions and Small Scale Structure}},\ }\href {https://doi.org/10.1016/j.physrep.2017.11.004} {\bibfield  {journal} {\bibinfo  {journal} {Phys. Rept.}\ }\textbf {\bibinfo {volume} {730}},\ \bibinfo {pages} {1} (\bibinfo {year} {2018})},\ \Eprint {https://arxiv.org/abs/1705.02358} {arXiv:1705.02358 [hep-ph]} \BibitemShut {NoStop}%
\bibitem [{\citenamefont {Zurek}(2009)}]{Zurek:2008qg}%
  \BibitemOpen
  \bibfield  {author} {\bibinfo {author} {\bibfnamefont {K.~M.}\ \bibnamefont {Zurek}},\ }\bibfield  {title} {\bibinfo {title} {{Multi-Component Dark Matter}},\ }\href {https://doi.org/10.1103/PhysRevD.79.115002} {\bibfield  {journal} {\bibinfo  {journal} {Phys. Rev. D}\ }\textbf {\bibinfo {volume} {79}},\ \bibinfo {pages} {115002} (\bibinfo {year} {2009})},\ \Eprint {https://arxiv.org/abs/0811.4429} {arXiv:0811.4429 [hep-ph]} \BibitemShut {NoStop}%
\bibitem [{\citenamefont {Fornengo}\ \emph {et~al.}(2011)\citenamefont {Fornengo}, \citenamefont {Panci},\ and\ \citenamefont {Regis}}]{Fornengo:2011sz}%
  \BibitemOpen
  \bibfield  {author} {\bibinfo {author} {\bibfnamefont {N.}~\bibnamefont {Fornengo}}, \bibinfo {author} {\bibfnamefont {P.}~\bibnamefont {Panci}},\ and\ \bibinfo {author} {\bibfnamefont {M.}~\bibnamefont {Regis}},\ }\bibfield  {title} {\bibinfo {title} {{Long-Range Forces in Direct Dark Matter Searches}},\ }\href {https://doi.org/10.1103/PhysRevD.84.115002} {\bibfield  {journal} {\bibinfo  {journal} {Phys. Rev. D}\ }\textbf {\bibinfo {volume} {84}},\ \bibinfo {pages} {115002} (\bibinfo {year} {2011})},\ \Eprint {https://arxiv.org/abs/1108.4661} {arXiv:1108.4661 [hep-ph]} \BibitemShut {NoStop}%
\bibitem [{\citenamefont {Lewin}\ and\ \citenamefont {Smith}(1996)}]{Lewin:1995rx}%
  \BibitemOpen
  \bibfield  {author} {\bibinfo {author} {\bibfnamefont {J.~D.}\ \bibnamefont {Lewin}}\ and\ \bibinfo {author} {\bibfnamefont {P.~F.}\ \bibnamefont {Smith}},\ }\bibfield  {title} {\bibinfo {title} {{Review of mathematics, numerical factors, and corrections for dark matter experiments based on elastic nuclear recoil}},\ }\href {https://doi.org/10.1016/S0927-6505(96)00047-3} {\bibfield  {journal} {\bibinfo  {journal} {Astropart. Phys.}\ }\textbf {\bibinfo {volume} {6}},\ \bibinfo {pages} {87} (\bibinfo {year} {1996})}\BibitemShut {NoStop}%
\bibitem [{\citenamefont {Menendez}\ \emph {et~al.}(2012)\citenamefont {Menendez}, \citenamefont {Gazit},\ and\ \citenamefont {Schwenk}}]{Menendez:2012tm}%
  \BibitemOpen
  \bibfield  {author} {\bibinfo {author} {\bibfnamefont {J.}~\bibnamefont {Menendez}}, \bibinfo {author} {\bibfnamefont {D.}~\bibnamefont {Gazit}},\ and\ \bibinfo {author} {\bibfnamefont {A.}~\bibnamefont {Schwenk}},\ }\bibfield  {title} {\bibinfo {title} {{Spin-dependent WIMP scattering off nuclei}},\ }\href {https://doi.org/10.1103/PhysRevD.86.103511} {\bibfield  {journal} {\bibinfo  {journal} {Phys. Rev. D}\ }\textbf {\bibinfo {volume} {86}},\ \bibinfo {pages} {103511} (\bibinfo {year} {2012})},\ \Eprint {https://arxiv.org/abs/1208.1094} {arXiv:1208.1094 [astro-ph.CO]} \BibitemShut {NoStop}%
\bibitem [{\citenamefont {Del~Nobile}\ \emph {et~al.}(2015)\citenamefont {Del~Nobile}, \citenamefont {Kaplinghat},\ and\ \citenamefont {Yu}}]{DelNobile:2015uua}%
  \BibitemOpen
  \bibfield  {author} {\bibinfo {author} {\bibfnamefont {E.}~\bibnamefont {Del~Nobile}}, \bibinfo {author} {\bibfnamefont {M.}~\bibnamefont {Kaplinghat}},\ and\ \bibinfo {author} {\bibfnamefont {H.-B.}\ \bibnamefont {Yu}},\ }\bibfield  {title} {\bibinfo {title} {{Direct Detection Signatures of Self-Interacting Dark Matter with a Light Mediator}},\ }\href {https://doi.org/10.1088/1475-7516/2015/10/055} {\bibfield  {journal} {\bibinfo  {journal} {JCAP}\ }\textbf {\bibinfo {volume} {10}},\ \bibinfo {pages} {055}},\ \Eprint {https://arxiv.org/abs/1507.04007} {arXiv:1507.04007 [hep-ph]} \BibitemShut {NoStop}%
\bibitem [{\citenamefont {Aprile}\ \emph {et~al.}(2019)\citenamefont {Aprile} \emph {et~al.}}]{XENON:2019gfn}%
  \BibitemOpen
  \bibfield  {author} {\bibinfo {author} {\bibfnamefont {E.}~\bibnamefont {Aprile}} \emph {et~al.} (\bibinfo {collaboration} {XENON}),\ }\bibfield  {title} {\bibinfo {title} {{Light Dark Matter Search with Ionization Signals in XENON1T}},\ }\href {https://doi.org/10.1103/PhysRevLett.123.251801} {\bibfield  {journal} {\bibinfo  {journal} {Phys. Rev. Lett.}\ }\textbf {\bibinfo {volume} {123}},\ \bibinfo {pages} {251801} (\bibinfo {year} {2019})},\ \Eprint {https://arxiv.org/abs/1907.11485} {arXiv:1907.11485 [hep-ex]} \BibitemShut {NoStop}%
\bibitem [{\citenamefont {Ren}\ \emph {et~al.}(2018)\citenamefont {Ren} \emph {et~al.}}]{PandaX-II:2018xpz}%
  \BibitemOpen
  \bibfield  {author} {\bibinfo {author} {\bibfnamefont {X.}~\bibnamefont {Ren}} \emph {et~al.} (\bibinfo {collaboration} {PandaX-II}),\ }\bibfield  {title} {\bibinfo {title} {{Constraining Dark Matter Models with a Light Mediator at the PandaX-II Experiment}},\ }\href {https://doi.org/10.1103/PhysRevLett.121.021304} {\bibfield  {journal} {\bibinfo  {journal} {Phys. Rev. Lett.}\ }\textbf {\bibinfo {volume} {121}},\ \bibinfo {pages} {021304} (\bibinfo {year} {2018})},\ \Eprint {https://arxiv.org/abs/1802.06912} {arXiv:1802.06912 [hep-ph]} \BibitemShut {NoStop}%
\bibitem [{\citenamefont {Li}\ \emph {et~al.}(2023)\citenamefont {Li} \emph {et~al.}}]{PandaX:2022xqx}%
  \BibitemOpen
  \bibfield  {author} {\bibinfo {author} {\bibfnamefont {S.}~\bibnamefont {Li}} \emph {et~al.} (\bibinfo {collaboration} {PandaX}),\ }\bibfield  {title} {\bibinfo {title} {{Search for Light Dark Matter with Ionization Signals in the PandaX-4T Experiment}},\ }\href {https://doi.org/10.1103/PhysRevLett.130.261001} {\bibfield  {journal} {\bibinfo  {journal} {Phys. Rev. Lett.}\ }\textbf {\bibinfo {volume} {130}},\ \bibinfo {pages} {261001} (\bibinfo {year} {2023})},\ \Eprint {https://arxiv.org/abs/2212.10067} {arXiv:2212.10067 [hep-ex]} \BibitemShut {NoStop}%
\bibitem [{\citenamefont {Chang}\ \emph {et~al.}(2010)\citenamefont {Chang}, \citenamefont {Pierce},\ and\ \citenamefont {Weiner}}]{Chang:2009yt}%
  \BibitemOpen
  \bibfield  {author} {\bibinfo {author} {\bibfnamefont {S.}~\bibnamefont {Chang}}, \bibinfo {author} {\bibfnamefont {A.}~\bibnamefont {Pierce}},\ and\ \bibinfo {author} {\bibfnamefont {N.}~\bibnamefont {Weiner}},\ }\bibfield  {title} {\bibinfo {title} {{Momentum Dependent Dark Matter Scattering}},\ }\href {https://doi.org/10.1088/1475-7516/2010/01/006} {\bibfield  {journal} {\bibinfo  {journal} {JCAP}\ }\textbf {\bibinfo {volume} {01}},\ \bibinfo {pages} {006}},\ \Eprint {https://arxiv.org/abs/0908.3192} {arXiv:0908.3192 [hep-ph]} \BibitemShut {NoStop}%
\bibitem [{\citenamefont {Foot}(2014)}]{Foot:2014mia}%
  \BibitemOpen
  \bibfield  {author} {\bibinfo {author} {\bibfnamefont {R.}~\bibnamefont {Foot}},\ }\bibfield  {title} {\bibinfo {title} {{Mirror dark matter: Cosmology, galaxy structure and direct detection}},\ }\href {https://doi.org/10.1142/S0217751X14300130} {\bibfield  {journal} {\bibinfo  {journal} {Int. J. Mod. Phys. A}\ }\textbf {\bibinfo {volume} {29}},\ \bibinfo {pages} {1430013} (\bibinfo {year} {2014})},\ \Eprint {https://arxiv.org/abs/1401.3965} {arXiv:1401.3965 [astro-ph.CO]} \BibitemShut {NoStop}%
\bibitem [{\citenamefont {Clarke}\ and\ \citenamefont {Foot}(2017)}]{Clarke:2016eac}%
  \BibitemOpen
  \bibfield  {author} {\bibinfo {author} {\bibfnamefont {J.~D.}\ \bibnamefont {Clarke}}\ and\ \bibinfo {author} {\bibfnamefont {R.}~\bibnamefont {Foot}},\ }\bibfield  {title} {\bibinfo {title} {{Mirror dark matter will be confirmed or excluded by XENON1T}},\ }\href {https://doi.org/10.1016/j.physletb.2016.12.047} {\bibfield  {journal} {\bibinfo  {journal} {Phys. Lett. B}\ }\textbf {\bibinfo {volume} {766}},\ \bibinfo {pages} {29} (\bibinfo {year} {2017})},\ \Eprint {https://arxiv.org/abs/1606.09063} {arXiv:1606.09063 [hep-ph]} \BibitemShut {NoStop}%
\bibitem [{\citenamefont {Aalbers}\ \emph {et~al.}(2023{\natexlab{a}})\citenamefont {Aalbers}, \citenamefont {Pelssers}, \citenamefont {Angevaare},\ and\ \citenamefont {Morå}}]{jelle_aalbers_2023_7636982}%
  \BibitemOpen
  \bibfield  {author} {\bibinfo {author} {\bibfnamefont {J.}~\bibnamefont {Aalbers}}, \bibinfo {author} {\bibfnamefont {B.}~\bibnamefont {Pelssers}}, \bibinfo {author} {\bibfnamefont {J.~R.}\ \bibnamefont {Angevaare}},\ and\ \bibinfo {author} {\bibfnamefont {K.~D.}\ \bibnamefont {Morå}},\ }\href {https://doi.org/10.5281/zenodo.7636982} {\bibinfo {title} {Jelleaalbers/wimprates: v0.5.0}} (\bibinfo {year} {2023}{\natexlab{a}})\BibitemShut {NoStop}%
\bibitem [{\citenamefont {Baxter}\ \emph {et~al.}(2021)\citenamefont {Baxter} \emph {et~al.}}]{Baxter:2021pqo}%
  \BibitemOpen
  \bibfield  {author} {\bibinfo {author} {\bibfnamefont {D.}~\bibnamefont {Baxter}} \emph {et~al.},\ }\bibfield  {title} {\bibinfo {title} {{Recommended conventions for reporting results from direct dark matter searches}},\ }\href {https://doi.org/10.1140/epjc/s10052-021-09655-y} {\bibfield  {journal} {\bibinfo  {journal} {Eur. Phys. J. C}\ }\textbf {\bibinfo {volume} {81}},\ \bibinfo {pages} {907} (\bibinfo {year} {2021})},\ \Eprint {https://arxiv.org/abs/2105.00599} {arXiv:2105.00599 [hep-ex]} \BibitemShut {NoStop}%
\bibitem [{\citenamefont {Aprile}\ \emph {et~al.}(2024{\natexlab{c}})\citenamefont {Aprile} \emph {et~al.}}]{xenonnt_ybe}%
  \BibitemOpen
  \bibfield  {author} {\bibinfo {author} {\bibfnamefont {E.}~\bibnamefont {Aprile}} \emph {et~al.},\ }\bibfield  {title} {\bibinfo {title} {{Low-Energy Nuclear Recoil Calibration of XENONnT with a $^{88}$YBe} photoneutron sourcec}} (\bibinfo {year} {2024}{\natexlab{c}}),\ \bibinfo {note} {in preparation}\BibitemShut {NoStop}%
\bibitem [{\citenamefont {Collar}(2013)}]{collar_applications_2013}%
  \BibitemOpen
  \bibfield  {author} {\bibinfo {author} {\bibfnamefont {J.~I.}\ \bibnamefont {Collar}},\ }\bibfield  {title} {\bibinfo {title} {{Applications of an $^{88}Y/Be$ photo-neutron calibration source to Dark Matter and Neutrino Experiments}},\ }\href {https://doi.org/10.1103/PhysRevLett.110.211101} {\bibfield  {journal} {\bibinfo  {journal} {Phys. Rev. Lett.}\ }\textbf {\bibinfo {volume} {110}},\ \bibinfo {pages} {211101} (\bibinfo {year} {2013})},\ \Eprint {https://arxiv.org/abs/1303.2686} {arXiv:1303.2686 [physics.ins-det]} \BibitemShut {NoStop}%
\bibitem [{\citenamefont {Aprile}\ \emph {et~al.}(2024{\natexlab{d}})\citenamefont {Aprile} \emph {et~al.}}]{XENON:2024xgd}%
  \BibitemOpen
  \bibfield  {author} {\bibinfo {author} {\bibfnamefont {E.}~\bibnamefont {Aprile}} \emph {et~al.} (\bibinfo {collaboration} {XENON}),\ }\bibfield  {title} {\bibinfo {title} {{XENONnT WIMP Search: Signal \& Background Modeling and Statistical Inference}},\ }\href@noop {} {\  (\bibinfo {year} {2024}{\natexlab{d}})},\ \Eprint {https://arxiv.org/abs/2406.13638} {arXiv:2406.13638 [physics.data-an]} \BibitemShut {NoStop}%
\bibitem [{\citenamefont {Szydagis}\ \emph {et~al.}(2011)\citenamefont {Szydagis}, \citenamefont {Barry}, \citenamefont {Kazkaz}, \citenamefont {Mock}, \citenamefont {Stolp}, \citenamefont {Sweany}, \citenamefont {Tripathi}, \citenamefont {Uvarov}, \citenamefont {Walsh},\ and\ \citenamefont {Woods}}]{szydagis_nest_2011}%
  \BibitemOpen
  \bibfield  {author} {\bibinfo {author} {\bibfnamefont {M.}~\bibnamefont {Szydagis}}, \bibinfo {author} {\bibfnamefont {N.}~\bibnamefont {Barry}}, \bibinfo {author} {\bibfnamefont {K.}~\bibnamefont {Kazkaz}}, \bibinfo {author} {\bibfnamefont {J.}~\bibnamefont {Mock}}, \bibinfo {author} {\bibfnamefont {D.}~\bibnamefont {Stolp}}, \bibinfo {author} {\bibfnamefont {M.}~\bibnamefont {Sweany}}, \bibinfo {author} {\bibfnamefont {M.}~\bibnamefont {Tripathi}}, \bibinfo {author} {\bibfnamefont {S.}~\bibnamefont {Uvarov}}, \bibinfo {author} {\bibfnamefont {N.}~\bibnamefont {Walsh}},\ and\ \bibinfo {author} {\bibfnamefont {M.}~\bibnamefont {Woods}},\ }\bibfield  {title} {\bibinfo {title} {{NEST}: a comprehensive model for scintillation yield in liquid xenon},\ }\href {https://doi.org/10.1088/1748-0221/6/10/P10002} {\bibfield  {journal} {\bibinfo  {journal} {J. Inst.}\ }\textbf {\bibinfo {volume} {6}},\ \bibinfo {pages} {P10002} (\bibinfo {year} {2011})}\BibitemShut {NoStop}%
\bibitem [{\citenamefont {Aprile}\ \emph {et~al.}(2024{\natexlab{e}})\citenamefont {Aprile} \emph {et~al.}}]{xenonnt_analysis1}%
  \BibitemOpen
  \bibfield  {author} {\bibinfo {author} {\bibfnamefont {E.}~\bibnamefont {Aprile}} \emph {et~al.},\ }\bibfield  {title} {\bibinfo {title} {{XENONnT Analysis: Signal Reconstruction, Calibration, and Event Selection}},\ }\Eprint {https://arxiv.org/abs/2409.08778} {arXiv:2409.08778 [physics.data-an]}  (\bibinfo {year} {2024}{\natexlab{e}})\BibitemShut {NoStop}%
\bibitem [{\citenamefont {Aharmim}\ \emph {et~al.}(2013)\citenamefont {Aharmim} \emph {et~al.}}]{SNO:2011hxd}%
  \BibitemOpen
  \bibfield  {author} {\bibinfo {author} {\bibfnamefont {B.}~\bibnamefont {Aharmim}} \emph {et~al.} (\bibinfo {collaboration} {SNO}),\ }\bibfield  {title} {\bibinfo {title} {{Combined Analysis of all Three Phases of Solar Neutrino Data from the Sudbury Neutrino Observatory}},\ }\href {https://doi.org/10.1103/PhysRevC.88.025501} {\bibfield  {journal} {\bibinfo  {journal} {Phys. Rev. C}\ }\textbf {\bibinfo {volume} {88}},\ \bibinfo {pages} {025501} (\bibinfo {year} {2013})},\ \Eprint {https://arxiv.org/abs/1109.0763} {arXiv:1109.0763 [nucl-ex]} \BibitemShut {NoStop}%
\bibitem [{\citenamefont {{XENON Collaboration}}(2024{\natexlab{b}})}]{fuse}%
  \BibitemOpen
  \bibfield  {author} {\bibinfo {author} {\bibnamefont {{XENON Collaboration}}},\ }\href {https://doi.org/10.5281/zenodo.11551366} {\bibinfo {title} {{XENONnT}/fuse: Refactor xenonnt epix and wfsim code}} (\bibinfo {year} {2024}{\natexlab{b}})\BibitemShut {NoStop}%
\bibitem [{\citenamefont {{XENON Collaboration}}(2024{\natexlab{c}})}]{axidence}%
  \BibitemOpen
  \bibfield  {author} {\bibinfo {author} {\bibnamefont {{XENON Collaboration}}},\ }\href {https://doi.org/10.5281/zenodo.12791105} {\bibinfo {title} {{XENONnT}/axidence: strax-based data-driven accidental coincidence background simulation and peak-level salting}} (\bibinfo {year} {2024}{\natexlab{c}})\BibitemShut {NoStop}%
\bibitem [{\citenamefont {Aprile}\ \emph {et~al.}(2021)\citenamefont {Aprile} \emph {et~al.}}]{XENON:2020gfr}%
  \BibitemOpen
  \bibfield  {author} {\bibinfo {author} {\bibfnamefont {E.}~\bibnamefont {Aprile}} \emph {et~al.} (\bibinfo {collaboration} {XENON}),\ }\bibfield  {title} {\bibinfo {title} {{Search for Coherent Elastic Scattering of Solar $^8$B Neutrinos in the XENON1T Dark Matter Experiment}},\ }\href {https://doi.org/10.1103/PhysRevLett.126.091301} {\bibfield  {journal} {\bibinfo  {journal} {Phys. Rev. Lett.}\ }\textbf {\bibinfo {volume} {126}},\ \bibinfo {pages} {091301} (\bibinfo {year} {2021})},\ \Eprint {https://arxiv.org/abs/2012.02846} {arXiv:2012.02846 [hep-ex]} \BibitemShut {NoStop}%
\bibitem [{\citenamefont {Bahcall}\ \emph {et~al.}(1996)\citenamefont {Bahcall}, \citenamefont {Lisi}, \citenamefont {Alburger}, \citenamefont {De~Braeckeleer}, \citenamefont {Freedman},\ and\ \citenamefont {Napolitano}}]{Bahcall:1996qv}%
  \BibitemOpen
  \bibfield  {author} {\bibinfo {author} {\bibfnamefont {J.~N.}\ \bibnamefont {Bahcall}}, \bibinfo {author} {\bibfnamefont {E.}~\bibnamefont {Lisi}}, \bibinfo {author} {\bibfnamefont {D.~E.}\ \bibnamefont {Alburger}}, \bibinfo {author} {\bibfnamefont {L.}~\bibnamefont {De~Braeckeleer}}, \bibinfo {author} {\bibfnamefont {S.~J.}\ \bibnamefont {Freedman}},\ and\ \bibinfo {author} {\bibfnamefont {J.}~\bibnamefont {Napolitano}},\ }\bibfield  {title} {\bibinfo {title} {{Standard neutrino spectrum from B-8 decay}},\ }\href {https://doi.org/10.1103/PhysRevC.54.411} {\bibfield  {journal} {\bibinfo  {journal} {Phys. Rev. C}\ }\textbf {\bibinfo {volume} {54}},\ \bibinfo {pages} {411} (\bibinfo {year} {1996})},\ \Eprint {https://arxiv.org/abs/nucl-th/9601044} {arXiv:nucl-th/9601044} \BibitemShut {NoStop}%
\bibitem [{\citenamefont {Faham}\ \emph {et~al.}(2015)\citenamefont {Faham}, \citenamefont {Gehman}, \citenamefont {Currie}, \citenamefont {Dobi}, \citenamefont {Sorensen},\ and\ \citenamefont {Gaitskell}}]{Faham:2015kqa}%
  \BibitemOpen
  \bibfield  {author} {\bibinfo {author} {\bibfnamefont {C.~H.}\ \bibnamefont {Faham}}, \bibinfo {author} {\bibfnamefont {V.~M.}\ \bibnamefont {Gehman}}, \bibinfo {author} {\bibfnamefont {A.}~\bibnamefont {Currie}}, \bibinfo {author} {\bibfnamefont {A.}~\bibnamefont {Dobi}}, \bibinfo {author} {\bibfnamefont {P.}~\bibnamefont {Sorensen}},\ and\ \bibinfo {author} {\bibfnamefont {R.~J.}\ \bibnamefont {Gaitskell}},\ }\bibfield  {title} {\bibinfo {title} {{Measurements of wavelength-dependent double photoelectron emission from single photons in VUV-sensitive photomultiplier tubes}},\ }\href {https://doi.org/10.1088/1748-0221/10/09/P09010} {\bibfield  {journal} {\bibinfo  {journal} {JINST}\ }\textbf {\bibinfo {volume} {10}}\bibfield  {number} {\bibinfo  {number} { (09)},\ \bibinfo {pages} {P09010}},\ }\Eprint {https://arxiv.org/abs/1506.08748} {arXiv:1506.08748 [physics.ins-det]} \BibitemShut {NoStop}%
\bibitem [{\citenamefont {Akerib}\ \emph {et~al.}(2021)\citenamefont {Akerib} \emph {et~al.}}]{Akerib:2021pfd}%
  \BibitemOpen
  \bibfield  {author} {\bibinfo {author} {\bibfnamefont {D.~S.}\ \bibnamefont {Akerib}} \emph {et~al.} (\bibinfo {collaboration} {LZ}),\ }\bibfield  {title} {\bibinfo {title} {{Enhancing the sensitivity of the LUX-ZEPLIN (LZ) dark matter experiment to low energy signals}},\ }\href@noop {} {\  (\bibinfo {year} {2021})},\ \Eprint {https://arxiv.org/abs/2101.08753} {arXiv:2101.08753 [astro-ph.IM]} \BibitemShut {NoStop}%
\bibitem [{\citenamefont {Sorensen}(2011)}]{sorensen_anisotropic_2011}%
  \BibitemOpen
  \bibfield  {author} {\bibinfo {author} {\bibfnamefont {P.}~\bibnamefont {Sorensen}},\ }\bibfield  {title} {\bibinfo {title} {{Anisotropic diffusion of electrons in liquid xenon with application to improving the sensitivity of direct dark matter searches}},\ }\href {https://doi.org/10.1016/j.nima.2011.01.089} {\bibfield  {journal} {\bibinfo  {journal} {Nucl. Instrum. Meth. A}\ }\textbf {\bibinfo {volume} {635}},\ \bibinfo {pages} {41} (\bibinfo {year} {2011})},\ \Eprint {https://arxiv.org/abs/1102.2865} {arXiv:1102.2865 [astro-ph.IM]} \BibitemShut {NoStop}%
\bibitem [{\citenamefont {Aalbers}\ \emph {et~al.}(2023{\natexlab{b}})\citenamefont {Aalbers} \emph {et~al.}}]{LZ:2022lsv}%
  \BibitemOpen
  \bibfield  {author} {\bibinfo {author} {\bibfnamefont {J.}~\bibnamefont {Aalbers}} \emph {et~al.} (\bibinfo {collaboration} {LZ}),\ }\bibfield  {title} {\bibinfo {title} {{First Dark Matter Search Results from the LUX-ZEPLIN (LZ) Experiment}},\ }\href {https://doi.org/10.1103/PhysRevLett.131.041002} {\bibfield  {journal} {\bibinfo  {journal} {Phys. Rev. Lett.}\ }\textbf {\bibinfo {volume} {131}},\ \bibinfo {pages} {041002} (\bibinfo {year} {2023}{\natexlab{b}})},\ \Eprint {https://arxiv.org/abs/2207.03764} {arXiv:2207.03764 [hep-ex]} \BibitemShut {NoStop}%
\bibitem [{\citenamefont {Ma}\ \emph {et~al.}(2023)\citenamefont {Ma} \emph {et~al.}}]{PandaX:2022aac}%
  \BibitemOpen
  \bibfield  {author} {\bibinfo {author} {\bibfnamefont {W.}~\bibnamefont {Ma}} \emph {et~al.} (\bibinfo {collaboration} {PandaX}),\ }\bibfield  {title} {\bibinfo {title} {{Search for Solar B8 Neutrinos in the PandaX-4T Experiment Using Neutrino-Nucleus Coherent Scattering}},\ }\href {https://doi.org/10.1103/PhysRevLett.130.021802} {\bibfield  {journal} {\bibinfo  {journal} {Phys. Rev. Lett.}\ }\textbf {\bibinfo {volume} {130}},\ \bibinfo {pages} {021802} (\bibinfo {year} {2023})},\ \Eprint {https://arxiv.org/abs/2207.04883} {arXiv:2207.04883 [hep-ex]} \BibitemShut {NoStop}%
\bibitem [{\citenamefont {Angloher}\ \emph {et~al.}(2016)\citenamefont {Angloher} \emph {et~al.}}]{Angloher:2016jsl}%
  \BibitemOpen
  \bibfield  {author} {\bibinfo {author} {\bibfnamefont {G.}~\bibnamefont {Angloher}} \emph {et~al.} (\bibinfo {collaboration} {CRESST}),\ }\bibfield  {title} {\bibinfo {title} {{Limits on momentum-dependent asymmetric dark matter with CRESST-II}},\ }\href {https://doi.org/10.1103/PhysRevLett.117.021303} {\bibfield  {journal} {\bibinfo  {journal} {Phys. Rev. Lett.}\ }\textbf {\bibinfo {volume} {117}},\ \bibinfo {pages} {021303} (\bibinfo {year} {2016})},\ \Eprint {https://arxiv.org/abs/1601.04447} {arXiv:1601.04447 [astro-ph.CO]} \BibitemShut {NoStop}%
\bibitem [{\citenamefont {Agnes}\ \emph {et~al.}(2023)\citenamefont {Agnes} \emph {et~al.}}]{DarkSide-50:2022qzh}%
  \BibitemOpen
  \bibfield  {author} {\bibinfo {author} {\bibfnamefont {P.}~\bibnamefont {Agnes}} \emph {et~al.} (\bibinfo {collaboration} {DarkSide-50}),\ }\bibfield  {title} {\bibinfo {title} {{Search for low-mass dark matter WIMPs with 12~ton-day exposure of DarkSide-50}},\ }\href {https://doi.org/10.1103/PhysRevD.107.063001} {\bibfield  {journal} {\bibinfo  {journal} {Phys. Rev. D}\ }\textbf {\bibinfo {volume} {107}},\ \bibinfo {pages} {063001} (\bibinfo {year} {2023})},\ \Eprint {https://arxiv.org/abs/2207.11966} {arXiv:2207.11966 [hep-ex]} \BibitemShut {NoStop}%
\bibitem [{\citenamefont {Huang}\ \emph {et~al.}(2023)\citenamefont {Huang} \emph {et~al.}}]{PandaX:2023xgl}%
  \BibitemOpen
  \bibfield  {author} {\bibinfo {author} {\bibfnamefont {D.}~\bibnamefont {Huang}} \emph {et~al.} (\bibinfo {collaboration} {PandaX}),\ }\bibfield  {title} {\bibinfo {title} {{Search for Dark-Matter\textendash{}Nucleon Interactions with a Dark Mediator in PandaX-4T}},\ }\href {https://doi.org/10.1103/PhysRevLett.131.191002} {\bibfield  {journal} {\bibinfo  {journal} {Phys. Rev. Lett.}\ }\textbf {\bibinfo {volume} {131}},\ \bibinfo {pages} {191002} (\bibinfo {year} {2023})},\ \Eprint {https://arxiv.org/abs/2308.01540} {arXiv:2308.01540 [hep-ex]} \BibitemShut {NoStop}%
\bibitem [{\citenamefont {Nobile}\ \emph {et~al.}(2015)\citenamefont {Nobile}, \citenamefont {Kaplinghat},\ and\ \citenamefont {Yu}}]{nobile_direct_2015}%
  \BibitemOpen
  \bibfield  {author} {\bibinfo {author} {\bibfnamefont {E.~D.}\ \bibnamefont {Nobile}}, \bibinfo {author} {\bibfnamefont {M.}~\bibnamefont {Kaplinghat}},\ and\ \bibinfo {author} {\bibfnamefont {H.-B.}\ \bibnamefont {Yu}},\ }\bibfield  {title} {\bibinfo {title} {Direct detection signatures of self-interacting dark matter with a light mediator},\ }\href {https://doi.org/10.1088/1475-7516/2015/10/055} {\bibfield  {journal} {\bibinfo  {journal} {J. Cosmol. Astropart. Phys.}\ }\textbf {\bibinfo {volume} {2015}}\bibinfo  {number} { (10)},\ \bibinfo {pages} {055}}\BibitemShut {NoStop}%
\bibitem [{\citenamefont {Klos}\ \emph {et~al.}(2013)\citenamefont {Klos}, \citenamefont {Menéndez}, \citenamefont {Gazit},\ and\ \citenamefont {Schwenk}}]{klos_large-scale_2013}%
  \BibitemOpen
\bibfield  {number} {  }\bibfield  {author} {\bibinfo {author} {\bibfnamefont {P.}~\bibnamefont {Klos}}, \bibinfo {author} {\bibfnamefont {J.}~\bibnamefont {Menéndez}}, \bibinfo {author} {\bibfnamefont {D.}~\bibnamefont {Gazit}},\ and\ \bibinfo {author} {\bibfnamefont {A.}~\bibnamefont {Schwenk}},\ }\bibfield  {title} {\bibinfo {title} {Large-scale nuclear structure calculations for spin-dependent {WIMP} scattering with chiral effective field theory currents},\ }\href {https://doi.org/10.1103/PhysRevD.88.083516} {\bibfield  {journal} {\bibinfo  {journal} {Phys. Rev. D}\ }\textbf {\bibinfo {volume} {88}},\ \bibinfo {pages} {083516} (\bibinfo {year} {2013})}\BibitemShut {NoStop}%
\bibitem [{\citenamefont {Aprile}\ \emph {et~al.}(2022)\citenamefont {Aprile} \emph {et~al.}}]{XENON:2022ltv}%
  \BibitemOpen
  \bibfield  {author} {\bibinfo {author} {\bibfnamefont {E.}~\bibnamefont {Aprile}} \emph {et~al.} (\bibinfo {collaboration} {XENON}),\ }\bibfield  {title} {\bibinfo {title} {{Search for New Physics in Electronic Recoil Data from XENONnT}},\ }\href {https://doi.org/10.1103/PhysRevLett.129.161805} {\bibfield  {journal} {\bibinfo  {journal} {Phys. Rev. Lett.}\ }\textbf {\bibinfo {volume} {129}},\ \bibinfo {pages} {161805} (\bibinfo {year} {2022})},\ \Eprint {https://arxiv.org/abs/2207.11330} {arXiv:2207.11330 [hep-ex]} \BibitemShut {NoStop}%
\bibitem [{\citenamefont {Aprile}\ \emph {et~al.}(2024{\natexlab{f}})\citenamefont {Aprile} \emph {et~al.}}]{XENON:2024qgt}%
  \BibitemOpen
  \bibfield  {author} {\bibinfo {author} {\bibfnamefont {E.}~\bibnamefont {Aprile}} \emph {et~al.} (\bibinfo {collaboration} {XENON}),\ }\bibfield  {title} {\bibinfo {title} {{XENONnT Analysis: Signal Reconstruction, Calibration and Event Selection}},\ }\href@noop {} {\  (\bibinfo {year} {2024}{\natexlab{f}})},\ \Eprint {https://arxiv.org/abs/2409.08778} {arXiv:2409.08778 [hep-ex]} \BibitemShut {NoStop}%
\bibitem [{\citenamefont {Feldman}\ and\ \citenamefont {Cousins}(1998)}]{Feldman:1997qc}%
  \BibitemOpen
  \bibfield  {author} {\bibinfo {author} {\bibfnamefont {G.~J.}\ \bibnamefont {Feldman}}\ and\ \bibinfo {author} {\bibfnamefont {R.~D.}\ \bibnamefont {Cousins}},\ }\bibfield  {title} {\bibinfo {title} {{A Unified approach to the classical statistical analysis of small signals}},\ }\href {https://doi.org/10.1103/PhysRevD.57.3873} {\bibfield  {journal} {\bibinfo  {journal} {Phys. Rev. D}\ }\textbf {\bibinfo {volume} {57}},\ \bibinfo {pages} {3873} (\bibinfo {year} {1998})},\ \Eprint {https://arxiv.org/abs/physics/9711021} {arXiv:physics/9711021} \BibitemShut {NoStop}%
\bibitem [{\citenamefont {{XENON Collaboration}}(2024{\natexlab{d}})}]{alea}%
  \BibitemOpen
  \bibfield  {author} {\bibinfo {author} {\bibnamefont {{XENON Collaboration}}},\ }\href {https://doi.org/10.5281/zenodo.10829030} {\bibinfo {title} {{XENONnT}/alea: A tool to perform toymc-based inference constructions}} (\bibinfo {year} {2024}{\natexlab{d}})\BibitemShut {NoStop}%
\bibitem [{\citenamefont {Cowan}\ \emph {et~al.}(2011)\citenamefont {Cowan}, \citenamefont {Cranmer}, \citenamefont {Gross},\ and\ \citenamefont {Vitells}}]{Cowan:2011an}%
  \BibitemOpen
  \bibfield  {author} {\bibinfo {author} {\bibfnamefont {G.}~\bibnamefont {Cowan}}, \bibinfo {author} {\bibfnamefont {K.}~\bibnamefont {Cranmer}}, \bibinfo {author} {\bibfnamefont {E.}~\bibnamefont {Gross}},\ and\ \bibinfo {author} {\bibfnamefont {O.}~\bibnamefont {Vitells}},\ }\bibfield  {title} {\bibinfo {title} {{Power-Constrained Limits}},\ }\href@noop {} {\  (\bibinfo {year} {2011})},\ \Eprint {https://arxiv.org/abs/1105.3166} {arXiv:1105.3166 [physics.data-an]} \BibitemShut {NoStop}%
\bibitem [{lig(2024)}]{lightwimp_data_release}%
  \BibitemOpen
  \href@noop {} {\bibinfo {title} {{XENONnT Light WIMP Data Release}}},\ \bibinfo {howpublished} {\url{https://github.com/XENONnT/light_wimp_data_release}} (\bibinfo {year} {2024})\BibitemShut {NoStop}%
\end{thebibliography}%

\end{document}